\documentclass[10pt,fullpage]{article}

\usepackage[includefoot,top = 1in, bottom=1in, right = 1in, left = 1in]{geometry}
\usepackage[T1]{fontenc}    
\usepackage{url}            
\usepackage{booktabs}       
\usepackage{lipsum}
\usepackage{graphicx}       
\usepackage{appendix}
\usepackage{mathptmx}   
\usepackage[linktocpage=true]{hyperref}
\hypersetup{
    colorlinks=true,
    linkcolor=blue,
    filecolor=magenta,      
    urlcolor=cyan,
    pdftitle={Overleaf Example},
    pdfpagemode=FullScreen,
}
\usepackage{geometry}
\usepackage{comment}
\usepackage{enumitem}

\setlist{nosep}
\usepackage{hyperref} 
\usepackage{longtable,ltcaption}
\usepackage{tocbibind}

\title{\textbf{Report on 2023 CyberTraining PI Meeting,\\
26-27 September 2023}}

\author{Geoffrey Fox, Univ. of Virginia (Chair), 
Mary Thomas, SDSC, UC San Diego (Report Lead) \\ 
\\
\textit{Program Committee Members: } \\
Sajal Bhatia, Sacred Heart Univ.), 
Marisa Brazil (Arizona State Univ.), \\
Nicole M Gasparini (Tulane Univ.),  
Venkatesh Mohan Merwade (Purdue Univ.), \\ 
Henry J. Neeman (Oklahoma Univ.)  \\  \\
\textit{Additional Contributors to Sessions and Report Preparation:}\\
Irfan Ahmed (Virginia Commonwealth Univ.), Jeff Carver (Univ. of Alabama), 
\\
Henri Casanova (Univ. of Hawaii),
Vipin Chaudhary (Case Western Reserve University),
\\
Dirk Colbry (Michigan State Univ.), 
Lonnie Crosby (Univ. of Tennessee Knoxville),  
\\Prasun Dewan (Univ. of North Carolina Chapel Hill), 
Jessica Eisma (The Univ. of Texas at Arlington), 
\\Nicole M Gasparini (Tulane Univ.), 
Kate Keahey (Argonne National Laboratory), 
\\ 
Qianqian Liu (Univ. of North Carolina Wilmington), 
Zhen Ni (Florida Atlantic Univ.),  \\
Sushil Prasad (Univ. of Texas at San Antonio), 
Apan Qasem (Texas State Univ.),  \\
Erik Saule (Univ. of North Carolina at Charlotte), 
Prabha Sundaravadivel (Univ. of Texas at Tyler),  \\
Karen Tomko (Ohio State Univ.)
}

\date{December 16, 2023}

\begin{document}
\pagenumbering{roman}

\maketitle

\begin{abstract}
This document describes a two-day meeting held for the Principal Investigators (PIs) of NSF CyberTraining grants. The report covers invited talks, panels, and six breakout sessions. The meeting involved over 80 PIs and NSF program managers (PMs). The lessons recorded in detail in the report are a wealth of information that could help current and future PIs, as well as NSF PMs, understand the future directions suggested by the PI community. The meeting was held simultaneously with that of the PIs of the NSF Cyberinfrastructure for Sustained Scientific Innovation (CSSI) program. This co-location led to two joint sessions: one with NSF speakers and the other on broader impact. Further, the joint poster and refreshment sessions benefited from the interactions between CSSI and CyberTraining PIs.
\end{abstract}

\newpage
\tableofcontents
\listoffigures
\listoftables
\newpage
\pagenumbering{arabic}

\section{Executive Summary }
\label{sec:exec-summary}

This report describes a two-day meeting held for the Principal Investigators (PIs) of NSF CyberTraining grants \cite{CyTrPIMtg23}. The report covers invited talks or panels and six breakout sessions. The meeting involved over 80 PIs and NSF program managers (PMs). The lessons recorded in detail in the report are a wealth of information that could help current and future PIs, as well as NSF PMs, understand the future directions suggested by the PI community. The meeting was held simultaneously with that of the PIs of the NSF Cyberinfrastructure for Sustained Scientific Innovation (CSSI) program \cite{CSSIPIMtg23}. This co-location led to two good joint sessions: one with NSF speakers and the other on broader impact. Further, the joint poster and refreshment sessions benefited from the interactions between CSSI and CyberTraining PIs. 

We can divide the activities into three groups: future directions, current status, and the features of a good CyberTraining project. Future directions can be seen in the report under two breakout sessions: the CyberTraining technology (B1C) and the transformative directions and new opportunities, including collaborations (B2C) sessions. The current status group included the NSF overview, the best practices breakout in CyberTraining lessons from projects (B1A), the success stories session, and the posters and lightning talks that are available for nearly all the projects.\cite{cytrpi-lightround1}\cite{cytrpi-lightround2} Three breakouts focused on the features of a great CyberTraining project. These were sustainability (B2B), metrics and evaluation (B2A), and approaches to broader impact and participation with a joint CSSI panel on broader impacts – Creating and supporting diverse teams for participants and community (B1B).

\subsection{Future Directions}
\label{sec:future}

Examining opportunities for the future, the PI community identified new CyberTraining domain areas, including the expected continued growth of AI and collaborations, which could be encouraged in solicitations. We are making slow progress in advancing historically underrepresented minorities in the STEM and STEAM areas. This could be addressed by pro-active CyberTraining partnerships, which could be mounted with encouragement in federal solicitations. Opportunities in technology for CyberTraining included both new ideas such as ChatGPT and related AI, which are being used for extracting data, improving writing style, finding programming bugs, and providing support in learning. The rapid evolution of these technologies produces support challenges, including scaling their use. Sharing of training and expertise should be encouraged. The growing importance of even more complex technologies, such as Science Foundation models, will only exacerbate the problem. We further need to address the resource needs for AI in education, which have clearly ballooned. Here, it has been estimated that the total computing power of the world's hyperscale clouds and supercomputers has increased very recently by around a factor of four (perhaps to around 200 exaflops) due to the growth in AI accelerators. Education resources need to grow by a similar factor. 

The opportunity for larger CyberTraining projects (centers and institutes) was stressed. As the field expands, there is a need for the coordination,  sustainability, and scalability that a center could bring. A center could also build shared resources such as a centralized, searchable platform and repository for CyberTraining material.

General discussions and the surveys reported later expressed support for the joint meeting with CSSI and recommended that it continue in later years. We also recommend that planning begin earlier than it did this year. That would allow the meeting to be held in the summer and other synergies to be explored, such as co-location with a conference.

\subsection{Current Status of CyberTraining}
\label{sec:current}

\noindent Best practices (B1A) identified that CyberTraining includes the use of cyberinfrastructure to train people and prepare them with emergent skills. Cyberinfrastructure can be high-performance computing equipment, computer analytic tools, machine learning libraries, and so on.  It is essential to consider data, which is very domain-dependent, as a critical part of the cyberinfrastructure. Some attendees thought that virtual training doesn't engage students well, while others felt that virtual learning is here to stay. A hybrid virtual and physical training format will be an option to bring more trainees together; meanwhile, new interactive teaching tools should be considered to engage students. Many PIs shared their experiences of evaluating the successful stories of students’ training, student presentations, professional conference showcases, course development, and others. This can involve a professional evaluator or be done more informally.

A rich set of experiences can be seen in the 59 slides and 55 posters prepared by the projects as well as the highlights in the seven success story presentations.  The latter not only overviewed their project but also discussed resources needed, the value of sharing, and particular challenges and successes they found. The community was interested in approaches to improving adoption and methods of keeping learners’ attention. Again, sustainability came up.

\subsection{Features of a Good CyberTraining Project}
\label{sec:good-ciproj}

\noindent The two broader impact sessions, B1B and the joint session, both discussed the important role of mentors and role models in successful outreach to underrepresented communities. Also, both stressed reaching out to community colleges, smaller universities, and organizations that promote and support these special communities.

The B2A session reviewed both a wide range of metrics and the different outcomes and impacts that could be examined. The simple registered user metric can be refined based on activity at different stages of the learning process, where participation tends to decline through the pipeline. More proactive metrics, such as surveys and interviews, are valuable even if their coverage is incomplete. The challenge of long-term outcomes is well known, but these are still valuable, and LinkedIn and publication tracking were discussed as one source for discovering participant futures. This needs funding after the project ends or through a separate dedicated center.

Sustainability (B2B) continues to be a challenge with CyberTraining programs. As well as establishing centers, it was suggested to take advantage of the NSF TIP directorate to extend support for existing programs for scalable and sustained implementation. This could involve industry partnerships or creating a centralized, searchable platform/repository for sharing CyberTraining material. Findings of the sustainability study included that partnerships with industry and non-profit organizations can be a viable pathway for sustainability as long as there is a value proposition; sustainability challenges vary by type of institution and target group, and there is a duplication of efforts and divergence of platforms used for the sharing of training material.

\begin{table}[ht]
\caption{Full Agenda: NSF CSSI and CyberTraining - Joint Workshop, 9/26/2023 - 9/27/2023}

\vspace{3pt}
\noindent\makebox[\textwidth]{
\begin{tabular}{p{0.75in} p{4.0in} } 
\toprule
\multicolumn{2}{l}{\textbf{DAY 1: 9/26/2023}} \\
\midrule
TIME & SESSION   \\ 
\midrule
7:30 - 8:15 & Registration \& Breakfast \\
8:15 - 9:00 & CyberTraining Lightning Talks \& Breakfast  \\
9:00 - 10:00 & CSSI and CyberTraining Joint Session: Opening Remarks (Joint)\\  
10:00 - 11:00 & Joint Poster Session 1 \& Coffee \\   
11:00 - 12:30 & CyberTraining Breakout Session 1 (60 minutes) plus report out (30 minutes) \\ 
& B1A: Best practices in CybertTraining Lessons from Projects \\
& B1B: Approaches to broader impact/participation plus Parallel Joint 
Panel with CSSI) \\
& B1C: Technology for CybertTraining, including ChatGPT and AI \\
12:30 - 2:00 & Networking Lunch (Joint Session) \\
2:00 - 3:30 & Breakout 2  (60 minutes) plus report out (30 minutes) \\
& B2A: CybertTraining Metrics and Outcomes \\
& B2B: Sustainability  \\
& B2B: The Future: Transformative Directions and New Opportunities \\ 
3:30 - 4:00 & Break  \\
4:00 - 5:00 & CyberTraining Success Stories \\
5:00 - 7:00 & Joint Reception \& Poster Session 2 \\
\midrule
\multicolumn{2}{l}{\textbf{DAY 2: 9/27/2023 }} \\
\midrule
7:30 - 8:15 & Registration \& Breakfast \\
8:15 - 9:00 & CyberTraining Lightning Talks \& Breakfast  \\
9:00 - 10:30  & Joint Session with CyberTraining: Broader Impacts – Creating and supporting diverse teams for participants and community \\
10:30 - 11:30  & Joint Poster Session 3 \\
11:30 - 12:30 & Panel Wrap Up \\
12:30 - 1:30 & Networking Lunch (Joint) \\
1:30 - 3:00 & CSSI \& CyberTraining: Team Collaboration, breakouts, and/or special interest table discussions/collaboration \\
\bottomrule
\end{tabular}
}
\label{tab:workshop-agenda}
\end{table}

\section{Workshop Organization and Planning}
\label{sec:workshop-org-summary}

The 2023 CyberTraining Principal Investigator (PI) meeting \cite{CyTrPIMtg23} was held in conjunction with the PI meeting for the 2023 NSF Cyberinfrastructure for Sustained Scientific Innovation (CSSI) 2023 NSF CSSI PI Meeting \cite{CSSIPIMtg23}
on September 26-27, 2023. Through the co-location of the two meetings, it was hoped that the communities would interact and that potentially new collaborations and projects would be an important outcome. Some statistics summarizing the CyberTraining meeting are highlighted below and described in Section \ref{sec:apx-surveys}: 
\begin{itemize}
    \item \# Registered attendees = 82: either PIs or NSF program managers, 12 in common with CSSI; 
    \item \# Unique organizations = 68
    \item \# Unique NSF awards represented = 67
    \item \# NSF Representatives = 9:  1 in common with CSSI. These were Ashok Srinivasan, Juan Jenny Li, Tom Gulbransen, Sharmistha Bagchi-Sen, Sheikh Ghafoor, May Yuan, Sharon Broude Geva, and also our thank you to both (not registered) Varun Chandola, (NSF CSSI Meeting lead) and Chaitan Baru.
\end{itemize}
\vspace{12pt}

\noindent The organizing and planning committee met virtually (via Zoom). The chair of the committee, Fox, also attended CSSI organizing meetings to ensure reasonable consistency of the programs of the two overlapping meetings. 
As part of our planning for the meeting, we conducted pre- and post-meeting surveys of the known CyberTraining PIs. The pre-meeting input was used for planning both the CyberTraining agenda and to help plan our CSSI interactions (see Section \ref{sec:apx-pre-survey}). The post-meeting survey was designed to be quick and to gather impact information (see Section \ref{sec:apx-post-survey}). Based on the impact results, it was felt that both surveys were productive and impactful. Gathering more demographic material from our attendees would be interesting in the future.

{\begin{figure*}[h]
\centering
\begin{minipage}[b]{.5\textwidth}
\centering
\includegraphics[width=.95\linewidth]{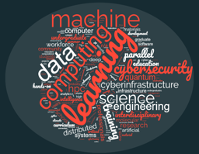}
\caption{Word cloud of keywords from submitted participants.} 
\label{fig:word-cloud-keywords}
\end{minipage} 
\hfill  
\begin{minipage}[b]{.48\textwidth}
\centering
\includegraphics[width= .95\linewidth]{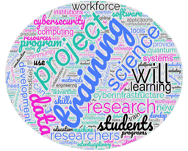}
\caption{Word cloud based on abstracts of submitted participant projects. } 
\label{fig:word-cloud-abstracts}
\end{minipage} 
\end{figure*}}
\vspace{.5em}

The scope of the projects is represented by the word cloud of keywords and abstracts in Figures \ref{fig:word-cloud-keywords} and \ref{fig:word-cloud-abstracts}. The original data for these clouds are given in the online resource 2023 NSF CyberTraining PI Meeting Base Material, which was gathered as part of the registration and survey processes (Appendix \ref{sec:apx-surveys}). Other materials contained in this resource are recorded in the appendices that list NSF project award numbers and titles (Appendix \ref{sec:attendee-list}), as well as submitted electronic lightning talks and posters (Appendix \ref{sec:presentations}).\cite{cytrpi-lightround1}\cite{cytrpi-lightround2} Note that a few posters and slides are missing when the participant is a collaborator and also for some awards released at the end of summer 2023. The two-day, joint meeting agenda can be found in Table \ref{tab:workshop-agenda}. In the workshop session sections below, we describe the lessons learned from particular sessions, where the structure was pre-planned by the overall committee, those leading the sessions, and comments from the surveys sent to PIs. Details about the workshop and planning can be found in Section \ref{sec:apx-wrkshop-plan}.

To estimate the impact of the meeting, we conducted a post-meeting survey.
As part of our planning for the meeting, we conducted pre- and post-meeting surveys of the known CyberTraining PIs. The pre-meeting input was used for planning both the CyberTraining agenda and to help plan our CSSI interactions (see Section \ref{sec:apx-pre-survey}). The post-meeting survey was designed to be quick and to gather impact information (see Section \ref{sec:apx-post-survey}). The overall response to the post-meeting survey was about 37\% (30/82 emails sent). The tables showing the data for the Meeting Usefulness and Participant Satisfaction survey data are described in detail in Section \ref{sec:apx-post-surv-usefulness}. Results indicate that, overall, attendees enjoyed the meeting and found it beneficial. For the different sessions, all received over 4.3+ average ratings. The poster and NSF participation were the most popular (about 4.7/5.0), while the lightning rounds were the least popular (4.23). The overall satisfaction for meeting logistics is a bit lower, 4.1, and this is mainly due to the location and hotel. In spite of this, the participants were overall satisfied. Finally, in response to whether or not participants enjoyed the co-location of the meeting with CSSI, the response was overwhelmingly yes (28/30) and just two neutral (see Section \ref{sec:apx-post-surv-cssi-colo}). 

Based on post-meeting survey results, the meeting and its associated sessions were productive and impactful.  Workshop resources, presentations, and participants lists are listed below, and details can be found in the Appendices: 
\begin{itemize}
    \item CyberTraining PI Meeting Website \cite{CyTrPIMtg23}
    \item List of Meeting Attendees: Table \ref{tab:attendee-list} 
    \item List of Presentations (Lightning \& Posters): Table \ref{tab:slides-posters}
    \item List of NSF Awards represented: Section \ref{sec:part-awards}
    \item Pre-meeting survey: Section \ref{sec:apx-pre-survey}
    \item Post-meeting survey: Section \ref{sec:apx-post-survey}
\end{itemize}

\section{Session Summaries}

\subsection{Day 1: NSF CSSI and CyberTraining Welcome (Joint Session)} 
\label{sec:day1-jointsession}

\begin{itemize}[label={}]
    \item  \textbf{Session Chairs:}   \textit{Christine Kirkpatrick \& Geoffrey Fox}
\end{itemize}
\vspace{12pt}

\noindent The meeting began with a joint session, including the CyberTraining PI and CSSI attendees meeting in the great hall. An overview of the meeting and its goals were presented by Christing Kirkpatrick (CSSI Meeting Chair) and Geoffrey Fox (CyberTraining PI Meeting Chair). The session included a welcoming talk by  Melissa Cragin (Rice University) and talks by NSF program officers. The agenda can be found in Table \ref{tab:day1-jointsession}.
\begin{table}[t]

\caption{Day 1 Agenda: NSF CSSI and CyberTraining - Joint Session}
\vspace{3pt}
\noindent\makebox[\textwidth]{
\begin{tabular}{p{0.70in}  p{1.25in} p{2.9in} p{.5in} } 
\toprule
TIME & PRESENTER & TOPIC & LINK  \\    [0.5ex] 
\midrule
9:00 - 9:10  & Christine Kirkpatric \& Geoffrey Fox & Opening Remarks   & 
\href{https://drive.google.com/file/d/1qEfoniYG9G2yOAsziKZS5el75icFtCZv/view?usp=drive_link}{[PDF]} \\  
9:10 - 9:20 & Melissa Cragin, Rice University & Welcome &  \\ 
9:20 - 9:30 & Katie Antypas & NSF Office of Advanced Cyberinfrastructure (OAC) &   \\ 
9:30 - 9:40 & Varun Chandola & NSF CI for Sustained Scientific Innovation (CSSI)   & \href{https://drive.google.com/file/d/13k00kFG6Q_a_btSrhkR1w5k_R3LF-uAv/view?usp=drive_link}{[PDF]} \\
9:40 - 9:50 & Ashok Srinivasan & NSF CyberTraining  & \href{https://drive.google.com/file/d/16roxJLeQvUxeJ7hLAcHdF5m13Epgyv43/view?usp=drive_link}{[PDF]} \\
9:50 - 10:00 & Chaitan Baru & NSF Directorate for Technology, Innovation, and Partnerships (TIP)   & \href{https://drive.google.com/file/d/1Ml7stOyl3k_FIoSSnr7UgICKTZCI8o89/view?usp=drive_link}{[PDF]} \\
 \bottomrule
\end{tabular}
}
\label{tab:day1-jointsession}
\end{table}

\subsection{B1A: Best practices in CyberTraining - Lessons from Projects }
\label{sec:b1a} 

\begin{itemize}[label={}]
    \item  \textbf{Session Chair:}   \textit{Dirk Colbry, MSU }
    \item  \textbf{Session CoChair:}   \textit{Mary Thomas, SDSC}
    \item  \textbf{Session Scribe:}   \textit{Zhen Ni, FAU }
\end{itemize}
\vspace{12pt}

In this interactive session, we asked the participants several questions:
(1)	What is CyberTraining? 
(2)	What motivated your current projects? 
(3)	What worked? 
(4)	What didn’t work?  
(5)	What do you use to evaluate success?
(6)	What do you need to know to write a CyberTraining grant?
Participants participated in open discussions and were asked to enter answers into a system called JamBoard, where each question is put up on a web page, and the audience can interact using a local web browser. The results are detailed in the Session B1A Details (Appendix \ref{sec:appdx-b1a})  below. 

The breakout session B1A was organized on Day 1 with 16+ attendees. There were five major questions during the session such as “What is CyberTraining?”, “What motivated your current projects?” “What worked?” “What has not worked?” “What do you use to evaluate success?” and “What do you need to know to write a CyberTraining grant?” Attendees used the Jamboard page to post their comments, and many also introduced their experiences verbally. The distinct notes are provided below. There are a few things to highlight. For example, attendees have consistent answers that CyberTraining includes the use of cyberinfrastructure to train people and prepare them with emergent skills. Cyberinfrastructure can be high-performance computing equipment, computer analytic tools, machine learning libraries, and so on.  Data is essential in this process, and data can be observed from different application domains. Some attendees have a strong motivation to bring machine learning techniques to marine science and provide students with background and knowledge so that students can apply machine learning methods for marine applications. While some attendees feel that virtual training doesn’t engage students well during the past several years, some PIs also show a sense that virtual learning may not go away. A hybrid format of virtual and physical training will be an option to bring more trainees together; meanwhile, new/interactive teaching tools should be considered to engage students. Besides a third-party evaluator, many PIs shared their experiences of evaluating the successful stories of students' training, student presentations, professional conference showcases, course development, and others.

\vspace{12pt}
\subsection{B1B: Approaches to broader impact and participation Breakout Session}
\label{sec:b1b}

\begin{itemize}[label={}]
    \item  \textbf{Session Chair:}   \textit{Jeff Carver, Alabama}
    \item  \textbf{Session CoChairs:}   \textit{Irfan Ahmed, Virginia Commenwealth, Nicole Gasparini, Tulane}
    \item  \textbf{Session Scribe:}   \textit{Marisa Brazil, ASU}
\end{itemize}
\vspace{12pt}

In this session, we discussed two “hopes” for increasing participation, along with a number of other general challenges and best practices. The first “hope” was to increase collaboration with community colleges as a mechanism for broadening participation. Some of the reasons for this approach include drawing from a more diverse student population and the fact that the first two years of higher education are critical to success. Some of the challenges with this idea include the competing priorities of the students, lack of connections with community colleges, and how to encourage community college students to take CyberTraining courses even if they were offered. The best practices included long-term collaborations, involving community college faculty in proposals, and working closely with community college instructors to develop the curriculum. The second “hope” was to increase the number of women and minority scientists in computing. The best practices for achieving this hope include providing role models, the presence of an inclusive environment (including diverse instructors), and using experiential learning. The key challenges are retaining students once they are there and identifying the best ways to measure impact. Finally, we discussed some more general challenges and best practices related to broadening participation. The general challenges included introducing CyberTraining at small Universities with a small number of faculty members, support for pre-tenure faculty who want to participate, budget for activities, and long-term sustainability. The best practice identified was to include CyberTraining in early courses that everyone must take and follow those with more specialized courses later in the curriculum.

\vspace{12pt}
\subsection{B1C: Technology for CyberTraining, including ChatGPT and AI}
\label{sec:b1c}

\begin{itemize}[label={}]
    \item  \textbf{Session Chair:}   \textit{Venkatesh Merwade, Purdue}
    \item  \textbf{Session CoChairs:}   \textit{Prasun Dewan UNC Chapel Hill, Jessica Eisma, UTA}
    \item  \textbf{Session Scribe:}   \textit{Henri Casanova, Hawaii}
\end{itemize}
\vspace{12pt}

\noindent This breakout session primarily focussed on four topics: (i) types of technologies used in CyberTraining, (ii) alternative systems or mechanisms used in CyberTraining, (iii) challenges associated with the use of technology in CyberTraining, and (iv) role of ChatGPT/AI in CyberTraining. Discussion from each topic is summarized below.

The CyberTraining community uses a range of software and hardware technologies. 

Perhaps the most fundamental of these is the programming environment trainees use for hands-on training.  Python and Jupyter Notebook emerged as the most commonly used programming language and platform by the community. The command line and GUI-based workflow systems were also used for data science training in some projects.

Tool-oriented training provides an opportunity to log trainee actions. Such logging has been used in some projects for objective evaluation of the training task that augments subjective survey-based evaluation. This evaluation has included metrics that identified how long trainees took on the problems and to what extent they solved the problems. Logging has also been explored by some projects to create visualizations that allow trainers to monitor the progress of trainees and intervene to help them. 

The technology that was most discussed was ChatGPT and the broader class of AI technologies in which it falls. Its potential to automate some of the trainers' actions by providing intelligent tutoring was discussed at length. AI, including ChatGPT is currently being used for extracting data, improving writing style, finding programming bugs, and providing support in learning. 

The community prefers the use of free, typically open-source, tools such as the ChatGPT UI, but commercial technologies, such as the ChatGPT API, are also used by a few to overcome some challenges. Some projects were developing their own tools - in particular, logging and visualization tools and tools that used the GPT-4 API directly.

There were some common issues raised by each of these technologies. Some were partially resolved, and some were unresolved.
The foremost among these challenges was how a  CyberTraining project should get access to and meaningfully use scalable and cost-effective versions of these technologies. For instance, how does a project get access to or create a scalable visualization or logging technology that can incrementally track trainees' progress? How does it make meaningful use of logging/visualization? Similarly, how does a project make meaningful use of the AI/ChatGPT technology (user interface or its API)  to provide training support?  ChatGPT is seen as a valuable tool, but there are some key issues related to its proper, cost-effective, real-time, ethical use that need to be addressed. 
Assuming the above problems can be addressed, another issue was what impact the use of each of these technologies has on the training given to the trainees. For example, what training has to be given to the trainees to use (a) a nonstandard, or even a standard, programming environment, (b) a technology that allows logging/visualization of their actions, and (c) AI/ChatGPT technology?

Many of these technologies can be used in multiple projects. How can common techniques and technologies be shared effectively by these projects? One project expressed willingness to share the logging capabilities it had developed for the Jupyter and Shell programming environments. A second project expressed its willingness to explore the use of the Jupyter logging capabilities. One of the participants in the group also attended an infrastructure session on sustainability, in which the session participants expressed excitement about joining infrastructure/CyberTraining projects to address such sharing.

Except for the last issue on technology sharing, these are mostly unresolved important issues that must be addressed to explore the comprehensive use of technology in CyberTraining.

\vspace{12pt}
\subsection{B2A: CyberTraining Metrics and Outcomes Breakout Session}
\label{sec:b2a}

\begin{itemize}[label={}]
    \item  \textbf{Session Chair:}   \textit{ Sushil Prasad, UTSA}
    \item  \textbf{Session CoChair:}   \textit{Lonnie Crosby, UTK}
    \item  \textbf{Session Scribe:}   \textit{Prabha Sundaravadivel, UT Tyler}
\end{itemize}
\vspace{12pt}

%

\noindent The session was organized under two main categories: (1) metrics and evaluation and (2) outcomes and impact.  Under “Metrics and Evaluation,” the following four subtopics were discussed: (a) important project metrics, (b) the role of surveys, (c) the role of learning objective-oriented assessment, and (d) evaluation metrics for indirect project assessments beyond surveys and exit quizzes.  Under “Outcomes and Impact,” the following three subtopics were discussed: (a) how do we measure the impact, (b) sharing educational materials digitally and incentives for such sharing, and (c) important broader impacts or other project outcomes.

This session focused on the collection of CyberTraining project metrics, project evaluation, project outcomes, and project impact. Beyond the usual CyberTraning project metrics such as number of attendees, downloads/views, or course adoptions - metrics were identified that constitute “leaky” pipelines where project engagement starts strong and decreases as the interactions become more substantial.  For example, looking at the progression between the number of course registrations, attendees, attendees who fully/actively participated, or attendees who completed assignments. These pipelines can give a direct measure of project engagement beyond tools such as surveys.  Additionally, these types of metrics were also mentioned in the context of indirect project assessments and broader impacts.  These metrics involved demographic information (i.e., participants' field of science or previous program participation), those that assess the quality of interactions (i.e., questions asked, attendance at office hours, or time spent on learning path), or those that assess the future impact (i.e., participants who pursue STEM, obtain internships, or curriculum adoption).  The use of social networks such as LinkedIn may provide methods to keep in touch with participants in order to collect data after their participation in the program.  Both the metrics and outcomes sessions mentioned that it may be useful to have project funding after the project ends in order to assess the impact. 

Evaluation mechanisms such as surveys, interviews, and publication tracking were observed to have some challenges, such as low response rates or implicit biases (surveys), scalability (interviews), or obtaining trackable citations (publications). Regardless, surveys were identified as being useful to evaluate satisfaction, identify improvement opportunities, and track knowledge before and after training. 

The sharing of CyberTraining materials was also discussed.  Challenges to sharing educational materials included the expected adoption of materials, the need for more effective sharing mechanisms, the lack of resources to maintain materials, and the lack of discoverability of these materials.   Potential solutions to these challenges were mentioned in the context of borrowing methods from the open-source community, such as the use of licenses for the identification of intellectual property, the use of citations (publications or digital-object-identifiers) to aid in discoverability, or utilizing mechanisms to receive feedback on materials via comments to aid in challenges with material maintenance. 

\vspace{12pt}
\subsection{B2B: Sustainability}
\label{sec:b2b}

\begin{itemize}[label={}]
    \item \textbf{Session Chair:} \textit{ Apan Qasem, Texas State Univ.} 
    \item \textbf{Session CoChair:} \textit{Karen Tomko, OSU}  
    \item \textbf{Session Scribe:} \textit{Marisa Brazil, ASU}  
\end{itemize}
\vspace{12pt}

\noindent The sustainability of CyberTraining Projects is Challenging as it is, in fact, for many NSF projects, including those in CSSI. The breakout session on sustainability examined both successful approaches and challenges. As in most NSF projects, it is hard to obtain funding to sustain a CyberTraining activity even though the costs can be minor, at least for the online educational resources. This funding could come from the Institution, Industry, or other NSF programs such as TIP. Not just funds are relevant; institutions should give credit in annual and tenure reviews for long term sustainable content. Sustainability is related to scalability, sharing, and diverse and broad community-building.. Central is the provision of a value proposition, clarification of distinctive features from the many other online resources,  and demonstration of impact. The structure of the CyberTraining project can use approaches that allow sustainability, such as virtual and asynchronous learning, clean packaging, and modular content. We need a central searchable Hub that will thoughtfully catalog projects and generate the users that will motivate continued effort and funding.

\vspace{12pt}
\subsection{B2C: The Future: Transformative Directions and New Opportunities Including Collaborations\label{sec:b2c}}

\begin{itemize}[label={}]
    \item  \textbf{Session Chair:}   \textit{Vipin Chaudhary, Case Western Reserve }
    \item  \textbf{Session CoChair:}   \textit{Geoffrey Fox, University of Virginia}
    \item  \textbf{Session Scribe:}   \textit{Erik Saule, UNC Charlotte}
\end{itemize}
\vspace{12pt}

\noindent The PI community identified new CyberTraining domain areas, including the expected continued growth of AI and collaborations, which could be encouraged in solicitations. We are making slow progress in advancing historically underrepresented minorities in the STEM and STEAM areas. This could be addressed by pro-active CyberTraining partnerships, which could be mounted with encouragement in federal solicitations. Opportunities in technology for CyberTraining included both new ideas such as ChatGPT and other LLM or Science Foundation models, plus addressing the resource needs for AI in education, which have clearly ballooned. Here, it has been estimated that the total computing power of the world's hyperscale clouds and supercomputers has increased very recently by around a factor of 4 (perhaps to 200 exaflops) due to the growth in AI accelerators. Education resources need to grow by a similar factor. The need for larger CyberTraining projects (centers and institutes) was stressed. As the field expands, there is a need for the coordination,  sustainability, and scalability that a center could bring.

\vspace{12pt}
\subsection{Success Stories Panel Session} 
\label{sec:success}

\begin{itemize}[label={}]
    \item  \textbf{Session Chair:}   \textit{Mary Thomas, SDSC }
    \item  \textbf{Session CoChair:}   \textit{Kate Keahey, ANL}
    \item  \textbf{Session Scribe:}   \textit{Qianqian Liu, UNC Wilmington}
\end{itemize}
\vspace{12pt}

\noindent The format of the success stories session was to hear from a group of projects about their work and outcomes. We asked each speaker to present a short talk in “lightning” talk form. we provided them with a set of slides and questions that we asked them to address: 
\begin{enumerate}
    \item Project Overview: What are the resource requirements of your training and education project?
    \item How do you find those resources?  What could be improved? 
    \item Do you share educational digital artifacts (e.g., software, homework assignments, etc.) related to your training and education projects? Could your projects benefit from artifacts produced by others?
    \item What have been your greatest challenges? 
    \item What are the key ingredients to being successful? 
\end{enumerate}
\vspace{12pt}

\vspace{12pt}
\noindent The Success Stories session consisted of 7 invited presentations, followed by audience Q\&A. Appendix \ref{sec:appdx-success} shows more details of the Q\&A session and includes Table \ref{tab:sucess-stories-presentations}, which has links to the presentation slides that are available for downloading. Several questions resulted in an active audience and panelist engagement, and these are highlighted below.
\\
\noindent \textbf{Question:} What are some best practices to facilitate the adoption of training materials/educational digital artifacts? \\
\noindent \textbf{Answer:} Putting the digital artifacts online and adding incentives (e.g., extra allocation, grant, or award) are needed to encourage adoption; we usually need extra committed support to organize workshops and to see the running products, identify industry support to facilitate adoption (need to be very careful); target your learners and ask for their needs -- if you can address their needs, they will go after your materials. 

\noindent \textbf{Question:} Is it true that if you build it, they will come? Are extra efforts needed to promote the materials? \\
\noindent \textbf{Answer: } Not true. A panelist (Dirk) described his successful program -- they have spent a lot of time building and selling the product. It is important to reach out to potential users. You need to let the learners know that what you developed meets their needs and can help their research. Be sure to minimize the hurdles for the learners, making the adoption easier for them. 

\noindent \textbf{Question:} How do we deal with the short attention span of the learners? \\
\noindent \textbf{Answer:} This question caused a lot of laughter among participants. But it is a serious issue. A Google search shows that the average human attention span on the web is 10-20 seconds. The group agreed that it takes much more time to teach HPC/CI concepts, so we have a big challenge. Suggestion by one panelist: use three sentences to capture your key points, and then move to three slides and to 1-hr or even 3-hr modules. Once the learners see why it is important, they will come back.

\noindent \textbf{Question:} How do you plan to run your projects after the money runs out? How do you make the project sustainable? \\
\noindent \textbf{Answer:} The suggestions include getting institutional support to host the materials. If young faculty members can get credit by maintaining the materials, they are more motivated to do this, go after more funding for long-term support, and ask for a small amount of yearly donation (like \$3) from end users.  

\noindent \textbf{Question:} Is it possible to make all these training modules and tutorials publicly available and share all the links? \\
\noindent \textbf{Answer:}  There is a repository which includes all NSF projects’ information. Also, there is the new HPC-ED project, which will build a federated repository where you can register information and pull information about other projects.

\subsection{Day 2: NSF CSSI and CyberTraining - Broader Impacts (Joint Session) }

\begin{itemize}[label={}]
    \item  \textbf{Session Chairs:}   \textit{Christine Kirkpatrick \& Geoffrey Fox }
\end{itemize}
\vspace{12pt}

\noindent The importance of broader impact was recognized in our plans, and the topic was covered in this joint session with CSSI on the second day. The theme of the session was \textit{NSF CSSI and CyberTraining Broader Impacts – Creating and supporting diverse teams and communities.} The session contained several significant presentations by leaders in the field (see Table \ref{tab:day2-speakers}). Presentation details and invited speaker bios can be found in the session details below. 

\begin{table}[ht]

\caption{Day 2 Agenda: NSF CSSI and CyberTraining - Joint Session}
\vspace{3pt}
\noindent\makebox[\textwidth]{
\begin{tabular}{p{0.8in}  p{1.05in} p{3.0in} p{.75in} } 
\toprule
TIME & PRESENTER & ORGANIZATION  & TALK \\ [0.5ex] 
\midrule
9:00 - 9:15 & Richard Alo & Florida A\&M University & [\href{https://drive.google.com/file/d/1fNh0sDWVR5liwspRR4X75ya-avFAH_Jr/view?usp=drive_link}{PDF}]   \\
9:15 - 9:30 & Linda Hayden & Elizabeth City State University, SGX3 & [\href{https://drive.google.com/file/d/1H0Vr4W2g7vmyZavo1lktU63KxN6kvNb2/view}{PDF}]    \\
9:30 - 9:45 & Dan Negrut & University of Wisconsin-Madison & [\href{https://drive.google.com/file/d/1U_bEHK1tyObnCS38tRhebVMjirijNJXi/view}{PDF}]   \\
9:45 - 10:00 & Sophie Kuchynka & Equity Accelerator & [\href{https://drive.google.com/file/d/1OxuASAvp67ySFcfyTcji2EctHqTO8_sI/view}{PDF}]    \\
10:00 - 10:15 & Reed Milewicz & Sandia National Laboratories & (n\/a) \\
10:15 - 10:30 & Frederi Viens & Rice University & [\href{https://drive.google.com/file/d/1WIy7FAXNMBdRWRIzfvsOG6NV7VSkLcWN/view}{PDF}]  \\
 \bottomrule
\end{tabular}
}
\label{tab:day2-speakers}
\end{table}

\textbf{Professor Richard Aló} presented on Creating and Supporting Diverse Teams and Communities. He is the Dean of the College of Science and Technology at FAMU and leads the NSF Florida Georgia Louis Stokes Alliance for Minority Participation FGLSAMP. His vision had three components: Reduce the Vast Underrepresentation in STEM (science, technology, engineering, and mathematics), Broaden Participation in STEAM (adding A for Arts), and Work Collaboratively to Equalize the Playing Field.  He gave alarming statistics on the university degrees at different levels granted to different ethnic groups (Asian, White,  HURM - historically under-represented minority). He stressed the value of virtual learning across the geographically broad FGLSAMP community. He described initiatives in data science and cybersecurity to conclude that CI and Data Science Provide Alternatives to Broadening Participation in STEAM. 

\textbf{Professor Linda Hayden} of Elizabeth City State University covered the results of a study by the NSF Office of Polar Programs Subcommittee on Diversity \& Inclusion. She stressed the uneven state of DEI efforts among different OPP awards. A highlight was the work of the NSF Science and Technology Center CReSIS. The REU program here involved women at the 42\% to 63\% level and minorities at the  66\% to 89\% level over a five-year period. She stressed the importance of partnerships rather than recruitment and recommended working with institutions with an underrepresented focus. So please partner with HBCU (Historically Black College and University), PBI (Predominantly Black Institution), MPO (Minority Professional Organization), or MSI (Minority Serving Institution). 

\textbf{Sophie Kuchynka} from the Equity Accelerator studied the role of the mentor-mentee relationship in three scenarios: high school, community colleges, and universities. The value of this was seen in all cases, but near-peer mentorship was particularly effective.

\section*{Acknowledgments}
This workshop was made possible with NSF Award \#2333991. We are grateful to NSF for helping to arrange the collaborative CyberTraining PI and CSSI PI meetings and thank the following NSF officers who attended the meeting: Ashok Srinivasan, Juan Jenny Li, Tom Gulbransen, Sharmistha Bagchi-Sen, Sheikh Ghafoor, May Yuan, Sharon Broude Geva, and also our thank you to Varun Chandola, (NSF CSSI Meeting lead). We would like to thank Christine Kirkpatrick and Lynne Schreiber (SDSC), for helping to coordinate the CSSI and CyberTraining PI Meetings. We also want to acknowledge the contributions of Benjamin Selwyn and Savanna Galambos (U.VA), for their work in creating and organizing the meeting website, registration and other logistics.

\bibliographystyle{unsrt}  
\bibliography{cytr-pi-mtg-2023}

\begin{thebibliography}{1}

\bibitem{CyTrPIMtg23}
NSF.
\newblock {CyberTraining Principal Investigator (PI) Meeting}.
\newblock \url{https://cybertrainingpi.wixsite.com/meeting2023}, 2023.
\newblock Accessed: November 3, 2023.

\bibitem{CSSIPIMtg23}
NSF.
\newblock {Cyberinfrastructure for Sustained Scientific Innovation (CSSI)
  Principal Investigator (PI) Meeting}.
\newblock \url{https://www.cssi-pi2023.org/}, 2023.

\bibitem{cytrpi-lightround1}
CyTrPI Committee.
\newblock Day 1 lightning talks.
\newblock
  \url{https://docs.google.com/presentation/d/18EDbDNjKkjQe0_25h_neAY_zdjuiwkXV/edit#slide=id.g283edc137e0_7_11},
  2023.
\newblock Accessed: November 30, 2023.

\bibitem{cytrpi-lightround2}
{CyTrPI Committee}.
\newblock Day 2 lightning talks.
\newblock
  \url{https://docs.google.com/presentation/d/1dFnlwGS8azlHEUI0FZNWkY-2KyvT_Lr7/edit#slide=id.g284034c7cc0_0_277},
  2023.

\bibitem{cytrpi-googlegroup}
{CyberTraining Principal Investigator Group}.
\newblock {CyberTraining Principal Investigator (PI) Google Group}.
\newblock \url{https://groups.google.com/u/1/g/cybertrainingpi2023}, 2023.
\newblock Accessed: December, 2023.

\bibitem{cytrpi-presurvey1}
{CyTrPI Program Committee}.
\newblock {Cybertraining PI Meeting Content Survey }.
\newblock
  \url{https://docs.google.com/forms/d/e/1FAIpQLScVilUK7tEzrp_CocEZ8xIEvrQCFnrbAaMELS3nj2FBjr2axQ/viewform?vc=0&c=0&w=1&flr=0},
  2023.
\newblock Accessed: December 16, 2023, 2023.

\bibitem{PostSurveyForm}
{CyTrPI Committee}.
\newblock {CyberTraining 2023 PI Meeting Feedback Survey}.
\newblock
  \url{https://docs.google.com/forms/d/1o4SoXHwCr1lrsShnRO3Chu3pYVfg2RGwiZWMuGw2KJo/edit},
  2023.

\bibitem{ACCESS}
ACCESS.
\newblock {Advanced Cyberinfrastructure Coordination Ecosystem: Services \&
  Support }.
\newblock \url{https://access-ci.org/}, 2023.
\newblock Accessed: November 1, 2023.

\end{thebibliography}


\appendix
\renewcommand{\thetable}{A\arabic{table}}

\section{Workshop Planning (Detailed Description)}
\label{sec:appdx-post-survey}
\label{sec:apx-wrkshop-plan} 

\subsection{Workshop Planning Overview}
\label{sec:apx-wrkshop-overview} 
\noindent 
The planning for the annual CyberTraining PI meeting was influenced heavily by the fact that the Program Officers for both the NSF CyberTraining (Ashok Srinivasan) and Cyberinfrastructure for Sustained Scientific Innovation (CSSI, Varun Chandola) requested that these programs hold a joint annual meeting.
It was hoped that, with a mix of shared and separate activities, the two programs would result in increased cross-project communications and impact and result in potentially new programs or proposals.

The CyberTraining PI meeting builds on previous PI meetings for other NSF programs. It provides a forum for: 
\begin{itemize}
    \item sharing technical information about projects with other colleagues, initiatives, and NSF program directors;
    \item exploring innovative topics related to the CyberTraining community; 
    \item discussing and learning about community practices;
    \item providing community feedback to NSF that could lead to new topics for CyberTraining solicitations;
    \item consider ways to ensure a diverse pipeline of CI researchers and professionals;
    \item investigate new ideas for achieving software and data sustainability.
\end{itemize}

At least one representative (PI/Co-PI/senior personnel) from each active CyberTraining project is required by NSF to attend the meeting and present a poster on their project. For a collaborative project with multiple awards, only one representative from the entire project is required. In addition, the CyberTraining PI (CyTrPI) Meeting program committee extended a special invitation to researchers from the Computational and Data-Enabled Science and Engineering (CDS\&E) and OAC Core Research (OAC Core) programs to participate in the meeting.

As part of our planning, we employed surveys to solicit ideas and suggestions from our PI community. The results of these surveys were used for input as we worked with the CSSI team. The results of the pre-workshop surveys helped shape the nature of the meeting (a mix of lightning Talks, Posters, Breakouts, and panels) and particular topics and issues within topics. Section \ref{sec:apx-surveys} contains a detailed description of the pre- and post-workshop surveys. The survey data was used to develop a more detailed program and to assess its impact.  
\begin{table}[h!]
\label{tab:session-leads}
\vspace{6pt}
\caption{CyberTraining PI Meeting: Session Leads}
\noindent\makebox[\textwidth]{
\begin{tabular}{|p{2.5in} | p{1.0in} |p{1.0in} |p{1.0in}  | } 
\toprule
SESSION & Chairs & Cochairs & Scribes  \\
\hline
Joint Session: CSSI and CyberTraining Opening Remarks (Joint)& Christine Kirkpatrick, SDSC &  Geoffrey Fox, UVA &   \\  
\hline
B1A: Best practices in CybertTraining Lessons from Projects  &  Dirk Colbry, MSU & Mary Thomas, SDSC & Zhen Ni, FAU \\
\hline
B1B: Approaches to broader impact/participation plus Parallel Joint Panel with CSSI &  Jeff Carver, Alabama & Irfan Ahmed, VCU, Nicole Gasparini, Tulane & Marisa Brazil, ASU  \\
\hline
B1C: Technology for CybertTraining, including ChatGPT and AI  &  Venkatesh Merwade, Purdue & Prasun Dewan, UNC Chapel Hill, Jessica Eisma, UTA & Henri Casanova, Hawaii \\
\hline
B2A: CybertTraining Metrics and Outcomes  &  Sushil Prasad, UTSA& Lonnie Crosby, UTK & Prabha Sundaravadivel, UTTyler \\
\hline
B2B: Sustainability   & Apan Qasem, Texas State & Karen Tomko, OSU & Marisa Brazil, ASU \\
\hline
B2C: The Future: Transformative Directions and New Opportunities  &  Vipin Chaudhary, Case Western & Geoffrey Fox, UVA& Erik Saule, UNC Charlotte \\ 
\hline
CyberTraining Success Stories  & Mary Thomas, SDSC  & Kate Keahey, ANL & Qianqian Liu, UNC-W \\
\hline
Joint Session: CSSI + CyberTraining: Broader Impacts – Creating and supporting diverse teams for participants and community  &  Christine Kirkpatrick, SDSC  & Geoffrey Fox, UVA &   \\
\bottomrule
\end{tabular}
}
\end{table}

\subsection{Organizing Team}
\label{sec:apx-org-team} 
The Organizing Team included the following Program Committee Members:
Chari: Geoffrey Fox, University of Virginia; and CoChairs: Sajal Bhatia, Sacred Heart University; Marisa Brazil, Arizona State University; Nicole M Gasparini, Tulane University; Venkatesh Mohan Merwade, Purdue University; Henry J. Neeman, Oklahoma University;  Mary Thomas, SDSC, UC San Diego. 

In addition, Benjamin Selwyn and Savanna Galambos (University of Virginia) contributed significantly in the areas of  organization, webstie, meals, registration, and other meeting logistics. We could not have completed the co-location of the CSSI and CyTrPI meetings without the help and support of CSSI team members Chrisine Kirkpatrick and Lynne Schreiber (SDSC).

Finally, the contributions and organization of all the breakout and joint sessions required significant effort. The session chairs, cochairs, and scribes are listed in Table \ref{tab:session-leads}.

\subsection{Meeting Attendee List}
\label{sec:attendee-list}

Overall, there were 74 participants, representing 67 NSF awards, of which, 16 registrants listed multiple NSF awards.
In addition, the following NSF officers also attended the meeting: 
Sharmistha Bagchi-Sen,  Chaitan Baru, Sharon Broude Geva,  Varun Chandola,
Sheikh Ghafoor, Tom Gulbransen, Jenny Li, Ashok Srinivasan, and May Yuan. 
Table \ref{tab:attendee-list} also shows a list of registered participants and the NSF awards they represent.

\fontsize{9}{11}\selectfont
\begin{longtable}{p{0.55in}  p{0.47in} p{0.45in} p{2.8in}  p{1.2in}} \caption{CyberTraining PI Attendee list} \\
\hline
LAST NAME	&	FIRST NAME	&	NSF AWARD	&	AWARD TITLE	&	ORGANIZATION	\\
\hline
Afflerbach	&	Benjamin	&	2017072	&	Collaborative Research: CyberTraining: Implementation: Medium: The Informatics Skunkworks Program for Undergraduate Research at the Interface of Data Science and Materials Science	&	University of Wisconsin - Madison	\\
Ahmed	&	Irfan	&	2017337	&	CyberTraining: Implementation: Small: Using Problem-Based Learning for Vocational Training in Cyberinfrastructure Security at Community Colleges	Virginia Commonwealth University	\\
Bagchi-Sen	&	Sharmistha	&	N/A	&	N/A	&	National Science Foundation	\\
Bhatia	&	Sajal	&	2017371	&	CyberTraining: Implementation: Small: Using Problem-Based Learning for Vocational Training in Cyberinfrastructure Security at Community Colleges	Sacred Heart University	\\
Bou--Harb	&	Elias	&	2230086	& Collaborative Research: CyberTraining: Implementation: Medium: Cross-Disciplinary Training for Joint Cyber-Physical Systems and IoT Security	&	Louisiana State University	\\
Brazil	&	Marisa	&	2230108	&	Collaborative Research: CyberTraining: CIP: A Cross-Institutional Research Engagement Network for CI Facilitators	&	Arizona State Universit\\
Brazil	&	Marisa	&	2230108	&	"
Cross--Institutional Research Engagement Network (CIREN)"	&	Arizona State University	\\
Carver	&	Jeff	&	2017259, 2017424	&	Collaborative Research: CyberTraining: Implementation: Small: INnovative Training Enabled by a Research Software Engineering Community of Trainers (INTERSECT)	&	University of Alabama	\\
Carver	&	Jeff	&	2017259, 2017424	&	Collaborative Research: CyberTraining: Implementation: Small: INnovative Training Enabled by a Research Software Engineering Community of Trainers (INTERSECT)	&	University of Alabama	\\
Casanova	&	Henri	&	1923621	&	Integrating core CI literacy and skills into university curricula via simulation-driven activities	&	University of Hawai`i at Manoa	\\
Chang	&	Philip	&	2229652	&	CyberTraining: Implementation: Small: CIberCATSS, A Comprehensive, Applied and Tangible CyberInfrastructure Summer School in Southeastern Wisconsin	University of Wisconsin-Milwaukee	\\
Chaudhary	&	Vipin	&	2320952	&	Collaborative Research: SCIPE: Interdisciplinary Research Support Community for Artificial Intelligence and Data Sciences	&	Case Western Reserve University	\\
Cleveland	&	Sean	&	2118222	&	Cyberinfrastructure training to Advance Environmental Science CI-TRACS Implementation: Medium Program	&	University of Hawaii - System	\\
Cleveland	&	Sean 	&	2118222	&	CyberTraining: Implementation: Medium: Cyberinfrastructure Training to Advance Environmental Science	&	University of Hawaii - System	\\
Colbry	&	Dirk	&	2118193, 1730137	&	CyberTraining: Pilot –A Professional Development and Certification Program (CCIFTD);
CyberTraining:CIP –Professional Skills for CyberAmbassadors"	&	Michigan State University	\\
Coles	&	Victoria	&	2321008	&	Collaborative Research: SCIPE: Enhancing the Transdisciplinary Research Ecosystem for Earth and Environmental Science with Dedicated Cyber Infrastructure Professionals	&	UMCES	\\
Coles	&	Victoria	&	2321008	&	gjh	&	UMCES	\\
Crichigno	&	Jorge 	&	2118311	&	CyberTraining on P4 Programmable Devices using an Online Scalable Platform with Physical and Virtual Switches and Real Protocol Stacks	&	University of South Carolina	\\
Cristea	&	Nicoleta	&	2117834	&	CyberTraining: Implementation: Medium: Machine Learning Training and Curriculum Development for Earth Science Studies	&	University of Washington	\\
Crosby	&	Lonnie	&	2230106, 2230108	&	Collaborative Research: CyberTraining: CIP: A Cross-Institutional Research Engagement Network for CI Facilitators	&	University of Tennessee, Knoxville	\\
Cui	&	Suxia	&	2321111	&	Collaborative Research: CyberTraining: Implementation: Small: Train the Trainers as Next Generation Leaders in Data Science for Cybersecurity for Underrepresented Communities	&	Prairie View A\&M University	\\
Dewan	&	Prasun	&	1924059  1829752	&	Collaborative Research: CyberTraining: Pilot: Semi-Automatic Assessment of Parallel Programs in Training of Students and Faculty ;    Collaborative Research: CyberTraining: CIU: Toward Distributed and Scalable Personalized Cyber-Training
	&	University of North Carolina	\\
Eisma	&	Jessica	&	2230054	&	CyberTraining: Pilot: Justice in Data: An intensive, mentored online bootcamp developing FAIR data competencies in undergraduate researchers in the water and energy sectors	&	University of Texas at Arlington	\\
Elmer	&	Peter	&	OAC-1829729	&	Collaborative Research: CyberTraining: CIC: Framework for Integrated Research Software Training in High Energy Physics (FIRST-HEP)	&	Princeton University	\\
Fox	&	Geoffrey	&	2200409	&	CyberTraining: CIC: CyberTraining for Students and Technologies from Generation Z	&	University of Virginia	\\
Ganapati	&	Sukumar	&	1924154	&	CyberTraining: Implementation: Medium: Advanced Cyber Infrastructure Training in Policy Informatics	&	Florida International University\\
Gasparini-she/her	&	Nicole	&	2118272	&	Collaborative Research: CyberTraining: Pilot: A CyberTraining Program to Advance Knowledge and Equity in the Geosciences	&	Tulane	\\
Ghafoor	&	Sheikh	&	12345	&	CyberTraining Program Director	&	National Science Foundation	\\
Ghafoor	&	Sheikh	&	1730417	&	CyberTraining: CDL: iPDC - Summer Institute for Integrating Parallel and Distributed Computing in Introductory Programming Classes	&	National Science Foundation	\\
Guan	&	Qiang	&	2230111	&	CyberTraining: Implementation: Small: Interactive and Integrated Training for Quantum Application Developers across Platforms	&	Computer Science, Kent State University	\\
Hayden 	&	Linda	&	2231406	&	SGX3 - A Center of Excellence to Extend Access, Expand the Community, and Exemplify Good Practices for CI Through Science Gateways.	&	SGX3/ECSU	\\
Hou	&	Daqing	&	2118079	&	CyberTraining: Pilot: Employing Proper Orthogonal Decomposition (POD) and High-Performance Computing (HPC) in Advanced CI	&	Clarkson University	\\
Jee	&	Kangkook	&	2321117	&	CyberTraining:Pilot:CyberTraining for Space CI in Low Earth Orbit (LEO)	&	The University of Texas at Dallas	\\
Jiang	&	Weiwen	&	2320957; 2311949	&	"CyberTraining: Pilot: Quantum Research Workforce Development on End-to-End Quantum Systems Integration;
Collaborative Research: OAC Core: An Integrated Framework for Enabling Temporal-Reliable Quantum Learning on NISQ-era Devices"	&	George Mason University	\\
Jiang	&	Weiwen	&	2320957; 2311949	&	CyberTraining: Pilot: Quantum Research Workforce Development on End-to-End Quantum Systems Integration; 
Collaborative Research: OAC Core: An Integrated Framework for Enabling Temporal-Reliable Quantum Learning on NISQ-era Devices	&	George Mason University	\\
Keahey	&	Kate	&	2230077	&	Collaborative Research: CyberTraining: Implementation: Medium: FOUNT: Scaffolded, Hands-On Learning for a Data-Centric Future	&	University of Chicago	\\
Knepper	&	Richard	&	2320977	&	CyberTraining: Pilot: HPC ED: Building a Federated Repository and Increasing Access through CyberTraining	&	Cornell University Center for Advanced Computing	\\
Knuth	&	Shelley	&	2138286	&	Track 2: Customized Multi-tier Assistance, Training, and Computational Help (MATCH) for End User ACCESS to CI	&	University of Colorado	\\
Kumar	&	Krishna	&	2321040	&	SCIPE: Chishiki.ai: A sustainable, diverse, and integrated CIP community for Artificial Intelligence in Civil and Environmental Engineering	&	University of Texas at Austin	\\
Liang	&	Xin	&	2311756	&	Collaborative Research: Elements: ProDM: Developing A Unified Progressive Data Management Library for Exascale Computational Science	&	University of Kentucky 	\\
Lindner	&	Peggy	&	2321110	&	Collaborative Research: CyberTraining: Implementation: Small: Train the Trainers as Next Generation Leaders in Data Science for Cybersecurity for Underrepresented Communities	&	University of Houston	\\
Liu	&	Qianqian	&	2230046	&	Collaborative Research: CyberTraining: Pilot: A CyberTraining Program to Advance Data Acquisition, Processing, and Machine Learning-based Modeling in Marine Science	&	University of North Carolina Wilmington	\\
Lou	&	Helen	&	2321055	&	CyberTraining: Pilot: Interdisciplinary Cybersecurity Education to Support Critical Energy and Chemical Infrastructure	&	Lamar University	\\
Lu	&	Xiaoyi	&	2321123	&	CyberTraining: Pilot: Cross-Layer Training of High-Performance Deep Learning Technologies and Applications for Research Workforce Development in Central Valley	&	University of California, Merced	\\
Martin	&	Thomas	&	2319979	&	Pilot: Machine Learning Foundations and Applications in the Earth Systems Sciences	&	Unidata	\\
Martin	&	Thomas	&	2319979	&	Pilot: Machine Learning Foundations and Applications in the Earth Systems Sciences	&	Unidata	\\
McHenry	&	Kenton	&	2230034, 2230035	&	Cyber2A: CyberTraining on AI-driven Analytics for Next Generation Arctic Scientists	&	University of Illinois Urbana-Champaign\\
Merwade	&	Venkatesh	&	2230092; 2230092	&	"Collaborative Research: CyberTraining: Implementation: Medium: Cyber Training for Open Science in Climate, Water and Environmental Sustainability;
CyberTraining: CIU:Cross-disciplinary Training for Findable, Accessible, Interoperable, and Reusable (FAIR) science"	&	Purdue University	\\
Miller	&	Adam	&	1829740	&	CyberTraining: CIU: The LSST Data Science Fellowship Program	&	Northwestern University/CIERA	\\
Mirkouei	&	Amin	&	2229604	&	Collaborative Research: CyberTraining: Implementation: Medium: CyberTraining of Construction (CyCon) Research Workforce Through an Educational and Community Engagement Platform	&	University of Idaho	\\
Ni	&	Zhen	&	1949921 1923983	&	Collaborative Research: CyberTraining: Implementation: Small: Multi-disciplinary Training of Learning, Optimization and Communications for Next Generation Power Engineers	&	Florida Atlantic University	\\
Nite	&	Sandra	&	1730695	&	Cyberinfrastructure Security Education for Professional and Students	&	Texas A\&M University	\\
Nomura	&	Ken-ichi	&	OAC-2118061	&	CyberMAGICS: Cyber Training on Materials Genome Innovation for Computational Software for Future Engineers	&	University of Southern California	\\
Prasad	&	Sushil	&	2017590 2002649	&	
Collaborative Research:CyberTraining:Implementation: Medium: Broadening Adoption of Parallel and Distributed Computing in Undergraduate Computer Science and Engineering Curricula; Collaborative Research: CyberTraining: Conceptualization: Planning a Sustainable Ecosystem for Incorporating Parallel and Distributed Computing into Undergraduate Education	&	University of Texas at San Antonio 	\\
Purwanto	&	Wirawan	&	2320998	&	Collaborative Research: CyberTraining: Implementation: Medium: T3-CIDERS: A Train-the-Trainer Approach to Fostering CI- and Data-Enabled Research in Cybersecurity	&	Old Dominion University	\\
Qasem	&	Apan	&	1829644	&	CyberTraining:CIP: Widening the CI Workforce On-ramp by Exposing Undergraduates to Heterogeneous Computing	&	Texas State University	\\
Rai	&	Neeraj	&	2118204	&	Collaborative Research: CyberTraining: Implementation: Medium: Establishing Sustainable Ecosystem for Computational Molecular Science Training and Education	&	Mississippi State University	\\
Roberts	&	Amy	&	2017760	&	CyberTraining: Implementation: Small: Enabling Dark Matter Discovery through Collaborative CyberTraining	&	CU Denver	\\
Samadi	&	Vidya 	&	2320979	&	Collaborative Research: CyberTraining: Implementation: Small: Inclusive Cyberinfrastructure and Machine Learning Training to Advance Water Science Research	&	Clemson University	\\
Saule	&	Erik	&	1924057	&	Aligning Learning Materials with Curriculum Standards to Integrate Parallel and Distributed Computing Topics in Early CS Education	&	UNC Charlotte	\\
Shah	&	Jindal	&	2118204  2118217  2200907  2118180  2118174	&	Collaborative Research: CyberTraining: Implementation: Medium: Establishing Sustainable Ecosystem for Computational Molecular Science Training and Education	&	Oklahoma State University	\\
Shook	&	Eric	&	1829708	&	Collaborative Research: CyberTraining: CIU: Hour of Cyberinfrastructure: Developing Cyber Literacy for Geographic Information Science	&	University of Minnesota	\\
Shu	&	Tong	&	2306184	&	Collaborative Research: CyberTraining: Pilot: Research Workforce Development for Deep Learning Systems in Advanced GPU Cyberinfrastructure	University of North Texas	\\
Sinkovits	&	Bob	&	2320934	&	CyberTraining: Implementation: Small: COMPrehensive Learning for end-users to Effectively utilize CyberinfraStructure (COMPLECS)	&	San Diego Supercomputer Center	\\
Snapp-Childs	&	Winona	&	2227627	&	RCN:CIP: Midwest Research Computing and Data Consortium	&	Indiana University	\\
Song	&	Houbing	&	2229975 2229976	&	Collaborative Research: CyberTraining: Pilot: Operationalizing AI/Machine Learning for Cybersecurity Training	&	University of Maryland, Baltimore County	\\
Song	&	Yang	&	2230046	&	Collaborative Research: CyberTraining: Pilot: A CyberTraining Program to Advance Data Acquisition, Processing, and Machine Learning-based Modeling in Marine Science	&	UNC Wilmington	\\
Srinivasan	&	Ashok	&	N/A	&	N/A	&	National Science Foundation	\\
Srivastava	&	Ankit	&	1829799	&	CyberTraining: CIC: The Texas A\&M University Computational Materials Science Summer School (CMS3)	&	Texas A\&M University	\\
Sundara vadivel	&	Prabha	&	1924117	&	Easy-Med: Interdisciplinary Training in Security, Privacy-Assured Internet of Medical Things	&	The University of Texas at Tyler\\
Takabi	&	Daniel	&	2118083	&	CyberTraining: Implementation: Small: Building Future Research Workforce in Trustworthy Artificial Intelligence (AI)	&	Old Dominion University\\
Thomas	&	Mary	&	2230127 2017767	&	CyberTraining CIP Training and Developing a Research Computing and Data CI Professionals (RCD-CIP) Community ; CyberTraining Implementation Small Developing a Best Practices Training Program in Cyberinfrastructure-Enabled Machine Learning Research &	San Diego Supercomputer Center, University of California San Diego	\\
Tomko	&	Karen	&	2320952, 2320953, 2320954	&	Collaborative Research: SCIPE: Interdisciplinary Research Support Community for Artificial Intelligence and Data Sciences	& Ohio Supercomputer Center	\\
Tomov	&	Stan	&	2017673	&	Linear Algebra Preparation for Emergent Neural Network Architectures (LAPENNA)	&	University of Tennessee, Knoxville	\\
Vaidyanathan	&	Ramachand ran (Vaidy)	&	2017233	&	Collaborative Research: CyberTraining: Implementation: Medium: Broadening Adoption of Parallel and Distributed Computing in Undergraduate Computer Science and Engineering Curricula	&	Louisiana State University	\\
Wang	&	Bing	&	2118102	&	CyberTraining: Pilot: Cyberinfrastructure Training in Computer Science and Geoscience	&	University of Connecticut	\\
Wong	&	Kwai	&	2017673	&	Linear Algebra Preparation for Emergent Neural Network Architectures (LAPENNA)	&	UTK, Knoxville	\\
Yang	&	Qing	&	2017564	&	CyberTraining: Implementation: Small: Collaborative and Integrated Training on Connected and Autonomous Vehicles Cyber Infrastructure	&	University of North Texas	\\
Yuan	&	Jiawei	&	2229976	&	Collaborative Research: CyberTraining: Pilot: Operationalizing AI/Machine Learning for Cybersecurity Training	&	University of Massachusetts Dartmouth	\\
Zhang	&	Jack (Yunpeng) 	&	2321110	&	Collaborative Research: CyberTraining: Implementation: Small: Train the Trainers as Next Generation Leaders in Data Science for Cybersecurity for Underrepresented Communities	&	University of Houston	\\
Zhang	&	Zhe	&	2321069	&	Collaborative Research: CyberTraining: Implementation: Small: Broadening Adoption of Cyberinfrastructure and Research Workforce Development for Disaster Management	&	Texas A\&M University 	\\
Zhou	&	Chi	&	2230025	&	CyberTraining: Implementation: Small: Infrastructure Cybersecurity Curriculum Development and Training for Advanced Manufacturing Research Workforce	University at Buffalo	\\
Zimmerman	&	Naupaka	&	2118305, 2118302	&	CyberTraining: Implementation: Medium: Collaborative Research: Computational and Data-Centric Ecology Training	&	University of San Francisco	\\
\bottomrule
\label{tab:attendee-list}
\end{longtable}
\normalsize

\subsubsection{Participant NSF Award List}
\label{sec:part-awards}

Below is a sorted list of the 67 unique NSF awards represented at the workshop:  \\

1730137, 1730695, 1829554, 1829644, 1829707, 1829708, 1829729, 1829740, 1829752, 1829764, 1829799, 1923621, 1923983, 1924057, 1924059, 1924117, 1924154, 1949921, 2002649, 2017072, 2017233, 2017259, 2017337, 2017371, 2017424, 2017564, 2017590, 2017673, 2017760, 2017767, 2117834, 2118061, 2118079, 2118083, 2118102, 2118174, 2118180, 2118193, 2118204, 2118217, 2118222, 2118272, 2118302, 2118305, 2118311, 2138286, 2200409, 2200907, 2227627, 2229603, 2229604, 2229652, 2229975, 2229976, 2230025, 2230034, 2230035, 2230046, 2230054, 2230077, 2230086, 2230092, 2230106, 2230108, 2230111, 2230127, 2231406

\subsection{Lightning Slide and Poster Presentations}
\label{sec:presentations}

The CyberTraining PI Meeting participants opted to hold daily presentations of their projects in the form of a "lightning round," where each PI presented one slide on their project. The talks were aggregated into two PDF files, which are listed below:
\begin{itemize}
    \item Day 1 Lightning Talks.pptx \href{https://docs.google.com/presentation/d/18EDbDNjKkjQe0\_25h\_neAY\_zdjuiwkXV/edit?usp=sharing&ouid=114300777188823967496&rtpof=true&sd=true}{[PPTX]}
    \item Day 2 Lightning Talks.pptx  \href{https://docs.google.com/presentation/d/1dFnlwGS8azlHEUI0FZNWkY-2KyvT_Lr7/edit?usp=sharing&ouid=114300777188823967496&rtpof=true&sd=true}{[PPTX]} 
\end{itemize}
\vspace{12pt}
\noindent Table \ref{tab:slides-posters} contains a listing of each participant who presented during the lighting rounds or the joint  posters sessions with CSSI. The table contains the presenters name, and the topic/title of the presentations, along with active URL links to the presentation (PPTX, PDF). 

\fontsize{9}{11}\selectfont
\begin{longtable}{p{0.58in}  p{0.5in} p{4.0in}  p{0.4in} p{0.4in} }
\caption{Listing of participants, lightning slides and posters} \\
\hline
LAST NAME	&	FIRST NAME	&	PROJECT TITLE	&	LIGHT-NING	&	POSTER \\
\midrule
Afflerbach	&	Benjamin	&	Collaborative Research: CyberTraining: Implementation: Medium: The Informatics Skunkworks Program for Undergraduate Research at the Interface of Data Science and Materials Science	&	\href{https://drive.google.com/open?id=1WsmueLa8Co-3DMH0_NOfdBn6mmsNLqF9}{[PPTX]}	&	\href{https://drive.google.com/open?id=11OCx0EXBSGCfF9JKmLKqCRhGomWd_DZj}{[PPTX]}  \\
Ahmed	&	Irfan	&	CyberTraining: Implementation: Small: Using Problem-Based Learning for Vocational Training in Cyberinfrastructure Security at Community Colleges	&	\href{https://drive.google.com/open?id=1-GvMunzXAt2jS9-I32eN-xp7Kt2BdJpb}{[PDF]}	&	\href{https://drive.google.com/open?id=1oW8OPUr5VnEsAbmWDwxOzz0GuSJVJ3dd}{[PDF]}  \\
Bhatia	&	Sajal	&	CyberTraining: Implementation: Small: Using Problem-Based Learning for Vocational Training in Cyberinfrastructure Security at Community Colleges	&	\href{ }{ }	&	\\
Bou-Harb	&	Elias	&	"
Collaborative Research: CyberTraining: Implementation: Medium: Cross-Disciplinary Training for Joint Cyber-Physical Systems and IoT Security"	&	\href{https://drive.google.com/open?id=1-FEPhY8BGyHD1AybkdzoNBuDWL7ZaS2E}{[PPTX]}	&	\href{https://drive.google.com/open?id=1WpOUbAHw3aGr8XvnFaz7XoYIxI1XEolr}{[PDF]}  \\
Brazil	&	Marisa	&	Collaborative Research: CyberTraining: CIP: A Cross-Institutional Research Engagement Network for CI Facilitators	&	\href{ }{ }	&	\href{ }{ }  \\

Carver	&	Jeff	&	Collaborative Research: CyberTraining: Implementation: Small: INnovative Training Enabled by a Research Software Engineering Community of Trainers (INTERSECT)	&	\href{https://drive.google.com/open?id=136wB2AFicVYLmJcswokOXeh8r2n9JCsn}{[PPTX]}	&	\href{https://drive.google.com/open?id=1GfTKsnOJ-NXioqYbdbOFh-QqM0pAOcNl}{[PDF]}  \\

Casanova	&	Henri	&	Integrating core CI literacy and skills into university curricula via simulation-driven activities	&	\href{https://drive.google.com/open?id=1zkeTadKwnt6v9b_2E53VKsI7YNRL0NyA}{[PDF]} & \href{https://drive.google.com/open?id=1Vl-JIOWQsrpMkKa2G66zQSjHL-Tpwv7O}{[PDF]}  \\		

Chang	&	Philip	&	CyberTraining: Implementation: Small: CIberCATSS, A Comprehensive, Applied and Tangible CyberInfrastructure Summer School in Southeastern Wisconsin	&	\href{https://drive.google.com/open?id=1aEnuiUUGS_ir8Qnuig8E4rhVQLDIZaUv}{[PPTX]}	&	\href{https://drive.google.com/open?id=1SXa4N8ZjlKMhYpqY9Xtx67HabJbR6S_2}{[PDF]}  \\
Chaudhary	&	Vipin	&	Collaborative Research: SCIPE: Interdisciplinary Research Support Community for Artificial Intelligence and Data Sciences	&	\href{https://drive.google.com/open?id=1b9oz9ox9IgA7rU1Rzz2rX1HYuPEu4XPH}{[PDF]} \href{https://drive.google.com/open?id=18bVMvZZlY3MkM99WjUGTchNIjAU8aO9D}{[PDF]}  \\	
Cleveland	&	Sean	&	CyberTraining: Implementation: Medium: Cyberinfrastructure Training to Advance Environmental Science	&	\href{https://drive.google.com/open?id=1_CDiFKaTpfx1NEvvrdZtZRsXyLNUhvQT}{[PPTX]}	&	\href{https://drive.google.com/open?id=1il0IaFX3ZYYEk82GEEYnirjqr4Zgt0tL}{[PPTX]}  \\
Colbry	&	Dirk	&	"CyberTraining:Pilot –A Professional Development and Certification Program (CCIFTD)
CyberTraining:CIP –Professional Skills for CyberAmbassadors"	&	\href{https://docs.google.com/presentation/d/1o9VlKWdJznUkG34fIpItANFSZ8jTNXsf/edit?usp=drive_link&ouid=114300777188823967496&sd=true}{[PPTX]}	&	\href{https://docs.google.com/presentation/d/1_EcFhV7agdbjX3HhI_pl4uE8qL4I0hAo/edit?usp=drive_link&ouid=114300777188823967496&rtpof=true&sd=true}{[PPTX]} \\
Coles	&	Victoria	&	Collaborative Research: SCIPE: Enhancing the Transdisciplinary Research Ecosystem for Earth and Environmental Science with Dedicated Cyber Infrastructure Professionals	&	\href{https://drive.google.com/open?id=1s7Ndc_K-KAYpUIoRP7oi5-0YO__OJUya}{[PDF]}  \\	
Crichigno	&	Jorge	&	CyberTraining on P4 Programmable Devices using an Online Scalable Platform with Physical and Virtual Switches and Real Protocol Stacks	&	\href{https://drive.google.com/open?id=1oS3VSupU7ks-Mm1OTMqXJR9_bxKuIbAp}{[PPTX]}	&	\href{https://drive.google.com/open?id=1U0pau7efapNgdMlp2GAieO0xrdV7IurE}{[PPTX]}  \\
Cristea	&	Nicoleta	&	CyberTraining: Implementation: Medium: Machine Learning Training and Curriculum Development for Earth Science Studies	&	\href{https://drive.google.com/open?id=1PvtBkWuiGPi92oHHpfr0pILrpYWx86jK}{[PPTX]}	&	\href{https://drive.google.com/open?id=1fGX67J-uL8GjZgEurvF3WvGEbXAOXX0e}{[PDF]}  \\
Crosby	&	Lonnie	&	Collaborative Research: CyberTraining: CIP: A Cross-Institutional Research Engagement Network for CI Facilitators	&	\href{https://drive.google.com/open?id=1O6xZUGoIHe9QMHnzFvoBC5Y7HrCL8TAq}{[PDF]}	&	\href{https://drive.google.com/open?id=1zKK4Qq9VuTqeE1AZRgnEsS1H5anioR1J}{[PDF]}  \\
Cui	&	Suxia	&	Collaborative Research: CyberTraining: Implementation: Small: Train the Trainers as Next Generation Leaders in Data Science for Cybersecurity for Underrepresented Communities	&	\href{ }{ }	&	\\

Dewan	&	Prasun	& Collaborative Research: CyberTraining: Pilot: Semi-Automatic Assessment of Parallel Programs in Training of Students and Faculty ;  Collaborative Research:CyberTraining: CIU: Toward Distributed and Scalable Personalized Cyber-Training &	\href{https://drive.google.com/open?id=1D9Eh2LGuS5SHLOfNb7jpgjQbp6W8jBp7}{[PDF]}	&	\href{https://drive.google.com/open?id=1_XGDO4EVsPQThFPHh6EOZHiTI9L_L5ee}{[PDF]}  \\

Eisma	&	Jessica	&	CyberTraining: Pilot: Justice in Data: An intensive, mentored online bootcamp developing FAIR data competencies in undergraduate researchers in the water and energy sectors	&	\href{https://drive.google.com/open?id=1pYznMPGQNKXkczvS_hWBCPDkUCUO2Ssi}{[PPTX]}	&	\href{https://drive.google.com/open?id=1ttR8NejbszyddiUpY5L1XzVBcLfP6mi7}{[PDF]}  \\
Elmer	&	Peter	&	Collaborative Research: CyberTraining: CIC: Framework for Integrated Research Software Training in High Energy Physics (FIRST-HEP)	&	\href{https://drive.google.com/open?id=1-FY0EF9dK6KRT3DZEMg0n8-JmTZEaNK_}{[PDF]}	&	\href{https://drive.google.com/open?id=1G5snZp6utAsDlyDyRKxZAZUGMbBBy4Xm}{[PDF]}  \\
Fox	&	Geoffrey	&	CyberTraining: CIC: CyberTraining for Students and Technologies from Generation Z	&	\href{https://drive.google.com/open?id=1vRPQmrZ8goyiwBKq-PtRWhYZNGydJj2_}{[PPTX]}	&	\href{https://drive.google.com/open?id=17NS408lhfbf9PsiGlsil_jvmkBrlNeC4}{[PPTX]}  \\
Ganapati	&	Sukumar	&	CyberTraining: Implementation: Medium: Advanced Cyber Infrastructure Training in Policy Informatics	&	\href{https://drive.google.com/open?id=1rDMK9rJdM6k-tze_kCwz2k7a6LbXh-0o}{[PPTX]}	&	\href{https://drive.google.com/open?id=1PB3dh5IomMzaY5rDTOdx2OyODW6JS0Sn}{[PDF]}  \\

Gasparini	&	Nicole	&	Collaborative Research: CyberTraining: Pilot: A CyberTraining Program to Advance Knowledge and Equity in the Geosciences	&	\href{https://drive.google.com/open?id=1gCDC-OMMy36XqJZIxMfz4Pr85pWVKTSO}{[PDF]}	&	\href{https://drive.google.com/open?id=1Ezq4uEeAbld1hlngVrZSk1gG8Hb3MfeI}{[PDF]}  \\

Guan	&	Qiang	&	CyberTraining: Implementation: Small: Interactive and Integrated Training for Quantum Application Developers across Platforms	&	\href{https://drive.google.com/open?id=1bqo3Nu0t2WMjQR152XhVlRaYkFf-GjRt}{[PPTX]}	&	\href{https://drive.google.com/open?id=1El1KsTCHW4TvfVoiAXcIdUyeGHAXrcOR}{[PDF]}  \\
Hayden	&	Linda	&	SGX3 - A Center of Excellence to Extend Access, Expand the Community, and Exemplify Good Practices for CI Through Science Gateways &	\href{https://drive.google.com/file/d/1H0Vr4W2g7vmyZavo1lktU63KxN6kvNb2/view}{PDF } \\

Hou	&	Daqing	&	CyberTraining: Pilot: Employing Proper Orthogonal Decomposition (POD) and High-Performance Computing (HPC) in Advanced CI	&	\href{ 118_Daqing Hou_PI-meeting-2023-lightning.PDF}{[PDF]}	&	\href{https://drive.google.com/open?id=1CgNDREaXV9dQ9R4g46NyCG6SiRTTx6LV}{[PPTX]}  \\
Jee	&	Kangkook	&	CyberTraining:Pilot:CyberTraining for Space CI in Low Earth Orbit (LEO)	&	\href{https://drive.google.com/open?id=1F7ANCRXwXa6QHMR0smHdk-cYXVLD_20B}{[PPTX]}	&	\href{https://drive.google.com/open?id=1if81tYo10K6wjpzMFaBVxRsgBPSoxjSh}{[PPTX]}  \\
Jiang	&	Weiwen	&	"CyberTraining: Pilot: Quantum Research Workforce Development on End-to-End Quantum Systems Integration

Collaborative Research: OAC Core: An Integrated Framework for Enabling Temporal-Reliable Quantum Learning on NISQ-era Devices"	&	\href{https://drive.google.com/open?id=1dPwuU4HAOAazqdOlHOZUGhYgyEJEdgy8}{[PPTX]}	&	\href{https://drive.google.com/open?id=18iBBj0z1itT5yeh1NqerRMn_wnCz55Ky}{[PDF]}  \\
Keahey	&	Kate	&	Collaborative Research: CyberTraining: Implementation: Medium: FOUNT: Scaffolded, Hands-On Learning for a Data-Centric Future	&	\href{https://drive.google.com/open?id=1Ep7tGwPl-VRvA7xwh3bkYre7NmcZoXWB}{[PPTX]}	&	\href{https://drive.google.com/open?id=1Fr33zOOX-Xj0GLvt3YmWgUm1LyyALWfZ}{[PDF]}  \\
Knepper	&	Richard	&	CyberTraining: Pilot: HPC ED: Building a Federated Repository and Increasing Access through CyberTraining	&	\href{https://drive.google.com/open?id=1WX3yTnCbm6YWds8BD8cJSZ3UZDMKemsx}{[PPTX]}	&	\href{https://drive.google.com/open?id=1RZS-qz8nJQ3VPUgh_VqH0oDjs_JQPvEd}{[PDF]}  \\
Knuth	&	Shelley	&	Track 2: Customized Multi-tier Assistance, Training, and Computational Help (MATCH) for End User ACCESS to CI	&	\href{https://docs.google.com/presentation/d/18SGZSSGJZLNLzxU1dtjuqkhLgHFaTK2P/edit?usp=drive_link&ouid=114300777188823967496&rtpof=true&sd=true}{[PPTX]}	&	\\
Kumar	&	Krishna	&	SCIPE: Chishiki.ai: A sustainable, diverse, and integrated CIP community for Artificial Intelligence in Civil and Environmental Engineering	&	\href{ }{ }	&	\\
Liang	&	Xin	&	Collaborative Research: Elements: ProDM: Developing A Unified Progressive Data Management Library for Exascale Computational Science	&	\href{ }{ }	&	\\
Lindner	&	Peggy	&	Collaborative Research: CyberTraining: Implementation: Small: Train the Trainers as Next Generation Leaders in Data Science for Cybersecurity for Underrepresented Communities	&	\href{https://drive.google.com/open?id=1Zs_DSCDPhQDvwA-Xx21_P5wHWdHXIu0b}{[PPTX]}	&	\\
Liu	&	Qianqian	&	Collaborative Research: CyberTraining: Pilot: A CyberTraining Program to Advance Data Acquisition, Processing, and Machine Learning-based Modeling in Marine Science	&	\href{https://drive.google.com/open?id=1PzNJve6HvfumUMWkL28MeLCeCSvLFcB8}{[PPTX]}	&	\href{https://drive.google.com/open?id=182Q80YdTzUQEXkDyES5IlJb58Kl0VELe}{[PPTX]}  \\
Lou	&	Helen	&	CyberTraining: Pilot: Interdisciplinary Cybersecurity Education to Support Critical Energy and Chemical Infrastructure	&	\href{https://drive.google.com/open?id=1sX3dN-TAvgBsrdCf_9sVABsNHCRxrGbD}{[PPTX]}	&	\href{https://drive.google.com/open?id=1-XHa-k61v5TuucHeKJZhqNq25MP6waf-}{[PDF]}  \\
Lu	&	Xiaoyi	&	CyberTraining: Pilot: Cross-Layer Training of High-Performance Deep Learning Technologies and Applications for Research Workforce Development in Central Valley	&	\href{https://drive.google.com/open?id=1yPrBf8LtvFRBqZy4hyiE_qNhTctHGttg}{[PDF]}	&	\\
Martin	&	Thomas	&	Pilot: Machine Learning Foundations and Applications in the Earth Systems Sciences	&	\href{https://drive.google.com/open?id=1KWYQj-Ofo1ycpr-RJfQUzSWGwYWAqAxi}{[PPTX]}	&	\href{https://drive.google.com/open?id=18gjuQqQqmGsm3UuozYFoER2IWx01dxA8}{[PDF]}  \\
McHenry	&	Kenton	&	Cyber2A: CyberTraining on AI-driven Analytics for Next Generation Arctic Scientists	&	\href{https://docs.google.com/presentation/d/1jh8xdAujrd2kygolQ6j226wqm4jRtD0v/edit?usp=drive_link&ouid=114300777188823967496&rtpof=true&sd=true}{[PPTX]}	&	\href{https://docs.google.com/presentation/d/1Ztj81-Swu4i0NMvxWcjHfW-UTRrMhATz/edit?usp=drive_link&ouid=114300777188823967496&rtpof=true&sd=true}{[PPTX]} \\

Merwade	&	Venkatesh	&	"Collaborative Research: CyberTraining: Implementation: Medium: Cyber Training for Open Science in Climate, Water and Environmental Sustainability

CyberTraining: CIU:Cross-disciplinary Training for Findable, Accessible, Interoperable, and Reusable (FAIR) science"	&	\href{https://drive.google.com/open?id=1lohxWLFXID9sZCo6C_HaquzYK75VpYKN}{[PDF]}	&	\href{https://drive.google.com/open?id=107h0jc47VqcSwdm5_H06hKdWWctcwI71}{[PDF]}  \\
Miller	&	Adam	&	CyberTraining: CIU: The LSST Data Science Fellowship Program	&	\href{https://drive.google.com/open?id=1YUZBk3VXGK7e9sZXH8nYxujTGDxXViWv}{[PDF]}	&	\href{https://drive.google.com/open?id=1WW9MFSiLcszIN2iNIKhY4rUmoxvHv40i}{[PDF]}  \\
Mirkouei	&	Amin	&	Collaborative Research: CyberTraining: Implementation: Medium: CyberTraining of Construction (CyCon) Research Workforce Through an Educational and Community Engagement Platform	&		&	\\
Ni	&	Zhen	&	Collaborative Research: CyberTraining: Implementation: Small: Multi-disciplinary Training of Learning, Optimization and Communications for Next Generation Power Engineers	&	\href{https://drive.google.com/open?id=1YtZnjQN8tgAPfuueR7YT0MLDMbpP1ePL}{[PPTX]}	&	\href{https://drive.google.com/open?id=1oZUFZUG8WNUuB0ofwF_5wYKWYczS4uMC}{[PPTX]}  \\
Nite	&	Sandra	&	Cyberinfrastructure Security Education for Professional and Students	&	\href{https://drive.google.com/open?id=1WTdw-PoThjQx7kya8D0rIAdBZbzwyyxL}{[PPTX]}	&	\href{https://drive.google.com/open?id=1rlhKVq9nTcfRF5MNEPFH3lVipXz198jg}{[PPTX]}  \\
Nomura	&	Ken-ichi	&	CyberMAGICS: Cyber Training on Materials Genome Innovation for Computational Software for Future Engineers	&	\href{https://drive.google.com/file/d/1g5vTkFjMFM0xKnrUGeRt4s9LRmMx-AGF/view?usp=drive_link}{[PDF]} 	&	\href{https://drive.google.com/file/d/1AGyX2yXyvMVXna2eFC-KGFtVKoUKao8g/view?usp=drive_link}{[PDF]} \\
Prasad	&	Sushil	&	Collaborative Research:CyberTraining:Implementation:Medium: Broadening Adoption of Parallel and Distributed Computing in Undergraduate Computer Science and Engineering Curricula; Collaborative Research: CyberTraining: Conceptualization: Planning a Sustainable Ecosystem for Incorporating Parallel and Distributed Computing into Undergraduate Education	&	\href{https://drive.google.com/open?id=1AafJAlSPz5SUj_YwuytVXDOX_OTGFCL3}{[PDF]}	&	\href{https://drive.google.com/open?id=1I78rZNFQIM6MUk22XZGh48IwhTFcdx0w}{[PDF]}  \\
Purwanto	&	Wirawan	&	Collaborative Research: CyberTraining: Implementation: Medium: T3-CIDERS: A Train-the-Trainer Approach to Fostering CI- and Data-Enabled Research in Cybersecurity	&	\href{https://drive.google.com/open?id=1u401HcKFfiYRcbaRJK_dIbHqoSW1fMOH}{[PPTX]}	&	\\
Qasem	&	Apan	&	CyberTraining:CIP: Widening the CI Workforce On-ramp by Exposing Undergraduates to Heterogeneous Computing	&	\href{https://drive.google.com/open?id=1kZeuJj5HbzwK38g_yZOOWlYOkm5aJWIw}{[PPTX]}	&	\href{https://drive.google.com/open?id=1gXPe3bdjfMQ684labNjX26Xyg_53FtaB}{[PDF]}  \\
Rai	&	Neeraj	&	Collaborative Research: CyberTraining: Implementation: Medium: Establishing Sustainable Ecosystem for Computational Molecular Science Training and Education	&	\href{https://drive.google.com/open?id=1FGfl_73jepxRXvjJ-b2_GeYtYr-8YQBw}{[PDF]}	&	\href{https://drive.google.com/open?id=1K_mdPRSW4cta9iu4jGk2E8FDCf9MJkHR}{[PDF]}  \\
Rashidi	&	Abbas	&	CyberTraining: Implementation: Small: Enabling Dark Matter Discovery through Collaborative CyberTraining	&	\href{https://drive.google.com/open?id=1RzjlRh95-cvmIzS6XRolCTvmsQoro6pR}{[PPTX]}	&	\href{https://drive.google.com/file/d/13jpk30SQVPY5DR_G5tLzI01s9I4uN-6X/view?usp=drive_link}{[PDF]}. \\
Roberts	&	Amy	&	Collaborative Research: CyberTraining: Implementation: Small: Inclusive Cyberinfrastructure and Machine Learning Training to Advance Water Science Research	&	\href{https://drive.google.com/open?id=1RrbAeNaP-GfU-0H-OSFNz0c4EoqSQFXk}{[PDF]}	&	\href{https://drive.google.com/open?id=1Hj6odhXVlHkPb7kgGCMgYVr-wG312ecl}{[PDF]}  \\
Samadi	&	Vidya	&	Aligning Learning Materials with Curriculum Standards to Integrate Parallel and Distributed Computing Topics in Early CS Education	&	2023 	& Award	\\
Saule	&	Erik	&	Collaborative Research: CyberTraining: Implementation: Medium: Establishing Sustainable Ecosystem for Computational Molecular Science Training and Education	&	\href{https://drive.google.com/open?id=1bBTTN0wh8r1CURLRqa80Xyx4yuzwpCk7}{[PDF]}	&	\href{https://drive.google.com/open?id=1bQ3nrG4u2KlS2vHtCibWLxjljiIRY_W6}{[PDF]}  \\
Shah	&	Jindal	&	Collaborative Research: CyberTraining: CIU: Hour of Cyberinfrastructure: Developing Cyber Literacy for Geographic Information Science	&	see Eric Shook	&	\\
Shook	&	Eric	&	Collaborative Research: CyberTraining: Pilot: Research Workforce Development for Deep Learning Systems in Advanced GPU Cyberinfrastructure	&	\href{https://drive.google.com/open?id=10LqTc1CHU1VxUPONINZE_ra1uAd9YT2S}{[PDF]}	&	\href{https://drive.google.com/open?id=1gVMnIX0rnwpwVhKMx92Y0ZvB1EsRS_Dc}{[PDF]}  \\
Shu	&	Tong	&	Collaborative Research: CyberTraining: Pilot: Research Workforce Development for Deep Learning Systems in Advanced GPU Cyberinfrastructure & \href{https://drive.google.com/file/d/1cIRo_C8eYmIx4YqafhKLPnpDW_PDE1uu/view}{[PDF]} & \href{https://drive.google.com/file/d/1Rjfcvg6R3JmW_XMJXKggE99JcInss709/view}{[PDF]} \\
Sinkovits	&	Bob	&	CyberTraining: Implementation: Small: COMPrehensive Learning for end-users to Effectively utilize CyberinfraStructure (COMPLECS)	&	\href{https://drive.google.com/open?id=1OxqiWBqMg6WBNPueZ-52Hmi7fj6cMx0d}{[PDF]}	&	\href{https://drive.google.com/open?id=1F0ioeSDdlpGTWMoxVwy7mBewfnmevNZ6}{[PDF]}  \\
Snapp-Childs	&	Winona	&	RCN:CIP: Midwest Research Computing and Data Consortium	&	\href{https://drive.google.com/open?id=11xO2AevSUlV90AR1IWMuQ4nazbCJXpho}{[PPTX]}	&	\href{https://drive.google.com/open?id=1z0dCGYtjuEUklrS1xzS1dNJGp_TXbKs7}{[PPTX]}  \\
Song	&	Houbing	&	Collaborative Research: CyberTraining: Pilot: Operationalizing AI/Machine Learning for Cybersecurity Training	&	\href{https://drive.google.com/open?id=1w0DLKy_M3ipkNtcJEcwICrDWpn7hnwR5}{[PPTX]}	&	\href{https://drive.google.com/open?id=1BEBDKHHU0ODTmVbYRuXNQOMKjjqvPurJ}{[PPTX]}  \\

Song	&	Yang	&	Collaborative Research: CyberTraining: Pilot: A CyberTraining Program to Advance Data Acquisition, Processing, and Machine Learning-based Modeling in Marine Science	&	\href{ }{ }	&	\href{ }{ }\\

Srivastava	&	Ankit	&	CyberTraining: CIC: The Texas A\&M University Computational Materials Science Summer School (CMS3)	&	\href{https://drive.google.com/open?id=1qHzaI6JVogtancwN5CDlJMP-ADWX_um4}{[PPTX]} & \href{https://drive.google.com/open?id=1JuGjN5ou0BTBiG9vF6idKQgH_lsQIdEE}{[PPTX]}  \\		

Sundarav-adivel	&	Prabha	&	Easy-Med: Interdisciplinary Training in Security, Privacy-Assured Internet of Medical Things	&	\href{https://drive.google.com/open?id=1ATtBmX0n3MUkNRnWyJmmLGcTwSvPvomd}{[PDF]} & \href{https://drive.google.com/open?id=1nRxo45BjUzDzFC3VI9-FuBXWCXpM12m5}{[PDF]}  \\		

Tomko	&	Karen	&	Collaborative Research: SCIPE: Interdisciplinary Research Support Community for Artificial Intelligence and Data Sciences	&	See V. 	&	Chaudhary  \\

Tomov	&	Stan	&	Linear Algebra Preparation for Emergent Neural Network Architectures (LAPENNA)	&	See Kwai	& Wong	\\

Vaidyanathan	&	 Vaidy	&	Collaborative Research: CyberTraining: Implementation: Medium: Broadening Adoption of Parallel and Distributed Computing in Undergraduate Computer Science and Engineering Curricula	&	\href{https://docs.google.com/presentation/d/1fpaUxeMFYyi2vD9IrgU2hGnrHV_5fXzN/edit?usp=drive_link&ouid=114300777188823967496&rtpof=true&sd=true}{[PPTX]}	&	\href{https://drive.google.com/open?id=1F9kOs7lEWfMNsZfPSFfyGPdw4s6AQLKU}{[PDF]}	  \\

Wang	&	Bing	&	CyberTraining: Pilot: Cyberinfrastructure Training in Computer Science and Geoscience	&	\href{https://drive.google.com/open?id=1WPa_TCFh61tpneTd35nJ-yR8Fcw4FuIf}{[PPTX]}	&	\href{https://drive.google.com/open?id=1PJKdW5_V92dujBII5qaYprMgU9CIWm0b}{[PPTX]}  \\

Wong	&	Kwai	&	Linear Algebra Preparation for Emergent Neural Network Architectures (LAPENNA)	&	\href{https://drive.google.com/open?id=1xeYn3z0nQnHxdQ2QhiQBPELKM-o-qEHX}{[PDF]}	&	\href{https://drive.google.com/open?id=1h11c0NLP9QEbPKLpx7dh_hC6JRwXez3w}{[PDF]}  \\

Yang	&	Qing	&	CyberTraining: Implementation: Small: Collaborative and Integrated Training on Connected and Autonomous Vehicles Cyber Infrastructure	&	\href{https://drive.google.com/open?id=1bT03bCi7zSBqPc2Pvn0SH430sLM5ZBno}{[PPTX]}	&	\href{https://drive.google.com/open?id=11hjx2RkmOKOpqwdMLhlPozxSI4Pp3u1m}{[PPTX]}  \\

Yuan	&	Jiawei	&	Collaborative Research: CyberTraining: Pilot: Operationalizing AI/Machine Learning for Cybersecurity Training	&	see  & Houbin Song 	\\

Zhang	&	Jack (Yunpeng)	&	Collaborative Research: CyberTraining: Implementation: Small: Train the Trainers as Next Generation Leaders in Data Science for Cybersecurity for Underrepresented Communities	&	2023 Award	&	\\

Zhang	&	Zhe	&	Collaborative Research: CyberTraining: Implementation: Small: Broadening Adoption of Cyberinfrastructure and Research Workforce Development for Disaster Management	&	n/a	&	\href{https://drive.google.com/open?id=1owJkSaUtUCsk-9JfkvdKLKXSE1f-TVgC}{[PPTX]}  \\

Zhou	&	Chi	&	CyberTraining: Implementation: Small: Infrastructure Cybersecurity Curriculum Development and Training for Advanced Manufacturing Research Workforce	&	\href{https://drive.google.com/open?id=1GU0Ot3eyD3RVNNhaxNcu1rLyi0TKA880}{[PPTX]}	&	\href{https://drive.google.com/open?id=1jjmZAHMpJttkhdV6HGKM-Z7ZNj6PKG5l}{[PDF]}  \\

Zimmerman	&	Naupaka	&	CyberTraining: Implementation: Medium: Collaborative Research: Computational and Data-Centric Ecology Training	&	\href{https://drive.google.com/open?id=1ysSGLd-nlFOLaoAdIf-WEctXL-4SYGoK}{[PDF]}	&	\href{https://drive.google.com/open?id=1hwez2LWwQv7eLFDDfYh8LUJE6LToOE3f}{[PDF]}  \\

\bottomrule
\label{tab:slides-posters}
\end{longtable}
\normalsize



\subsection{Workshop Surveys}
\label{sec:apx-surveys} 

As part of our planning for the meeting, we created a Google Group \cite{cytrpi-googlegroup} for communication and conducted pre- and post-meeting surveys of the known CyberTraining PIs (see Sections \ref{sec:apx-pre-survey} and \ref{sec:apx-post-survey}). The pre-meeting input was used for planning both the CyberTraining agenda and to help plan our CSSI interactions. The post-meeting survey was designed to be quick and to gather impact information. Based on the impact results, it was felt that both surveys were productive and impactful. In the future, gathering more demographic material from our attendees would be interesting.

The types of questions included usefulness and satisfaction. For the pre-meeting surveys, we asked participants to rate the options from most favorite to least favorite (e.g., 1-7, 1,0,-1) shown in Table \ref{tab:likert1}. For the post-workshop surveys, usefulness and satisfaction questions were rated using a Likert scale of 1:5, which is shown in Table \ref{tab:likert2} below.

\begin{table}[h]
\caption{Survey Ranking/Preference scales}

\vspace{6pt}
\noindent\makebox[\textwidth]{
\begin{tabular}{p{1.0in} p{0.5in} | p{1.5in}  p{0.5in} } 
\toprule
TOPIC RANK & VALUE & FORMAT PREFERENCE & VALUE \\ [0.5ex] 
\midrule
Most Preferred &  1  &  Like  &  1 \\
Neutral &  3 &  Maybe  & 0 \\
Least Preferred &  7 &  Dislike  &  -1 \\
\hline
\end{tabular}
}
\label{tab:likert1}
\end{table}

\begin{table}[h]
\caption{Likert satisfaction and usefulness scales}
\vspace{6pt}

\noindent\makebox[\textwidth]{
\begin{tabular}{p{1.0in} p{0.5in} | p{2.0in}  p{0.5in} } 
\toprule
USEFULNESS & VALUE & SATISFACTION RANKING & VALUE \\ [0.5ex] 
\midrule
Very useful &  5  &  Very satisfied  &  5 \\
Somewhat useful &  4 &  Satisfied & 4 \\
Neutral &  3 &  Neither satisfied nor dissatisfied  & 3 \\
A little useful &  2 &  Dissatisfied  &   2 \\
Not at all useful &  1 &  Very dissatisfied  &  1 \\
\hline
\end{tabular}
}
\label{tab:likert2}
\end{table}

\subsubsection{Pre-Workshop Survey}
\label{sec:apx-pre-survey} 

The results of the pre-workshop surveys helped shape the nature of the meeting (a mix of lightning Talks, Posters, Breakouts, and panels) and particular topics and issues within topics. \cite{cytrpi-presurvey1}

Table \ref{tab:survey-pre-content}  contains responses to the first survey on what content and session formats would be of interest to the CyberTraining PI community. The survey was sent out to approximately 90 PIs, with 34 responses, for an overall response rate of approximately 38\%. In the first survey, we asked respondents to rank seven topics for panels and breakouts (Table \ref{tab:survey-pre-content}) and their preferred formats (Table \ref{tab:survey-pre-format}). In a second follow-up survey, we asked respondents to list subtopic details for each of the seven topics with their ranking from the first survey in the second column. This input was used to develop a more detailed program, including a charge to breakout sessions and speakers for the Success Story session. The results of these surveys were used for input as we worked with the CSSI team. The final agenda (Table \ref{tab:workshop-agenda}) is the result of our planning, and the surveys were very helpful.

\begin{table}[ht]
\vspace{6pt}
\caption{Pre-Workshop Content Survey (lower is better)}
\begin{tabular}{p{4.5in} p{1.0in} } 
\toprule
QUESTION & RESPONSE  \\
\midrule
Best practices in CyberTraining  &  1.79 \\
CyberTraining metrics and outcomes, identifying what succeeded and what was less successful  &   2.82 \\
Success Stories  &   3.09 \\
Approaches to broader impact/participation  &   3.21 \\
ChatGPT and Other advanced AI  &   3.91 \\
Sustainability (CSSI choice) & 4.09 \\
Gaps in Coverage and collaboration (with CSSI or more generally) opportunity]   &  4.35 \\
\bottomrule
\end{tabular}
\label{tab:survey-pre-content}
\end{table}

Once the meeting content was determined, we asked about presentation formats in a follow-on survey, as shown in Table \ref{tab:survey-pre-format}. Based on the preferences of our community, we selected lightning talks, breakout sessions, and posters for the primary session organization. In some of the joint CSSI sessions and for the success stories session, we employed panels where it made sense.


\begin{table}[ht]
\vspace{6pt}
\caption{Pre-Workshop Presentation Format Preference (higher is better)}
\begin{tabular}{p{4.5in} p{1.0in} } 
\toprule
TOPIC & RESPONSE  \\
\midrule
I like Lightning talks &  0.68 \\
I like splitting into breakout sessions &  0.68 \\
I like Posters &  0.59 \\
I like Panels &  0.54 \\
I like invited talks & 0.44 \\
\bottomrule
\end{tabular}
\label{tab:survey-pre-format}
\end{table}

\vspace{12pt}
\noindent \textbf{Free Form Comments to the question: What do you want to get out of the meeting?} Note that this is some of the data used in the word clouds of keywords and abstract content shown in Figures \ref{fig:word-cloud-abstracts} and \ref{fig:word-cloud-keywords}.
\begin{itemize}
    \item I would prefer not to go. But in all seriousness, I think an opportunity to talk to other CyberTraining people and find common grounds for synergy might be super useful for me. I would particularly welcome meeting other CyberTraining people who want to collaborate on a larger version of CyberTraining.
    \item Learn from other projects and new resources at NSF to improve our project.
    \item To network with others
    \item Opportunities to establish new collaborations
    \item I'd like to learn more about what other groups are doing in terms of training students with little to no programming experience and training approaches to working with students with diverse domain science backgrounds
    \item Best practices from other projects, an understanding of where funding priorities are going.
    \item Hear talks about various projects rather than see just posters. Based on past experience, I dislike panels as they become a bunch of talks usually. I would rather see a broader set of talks of at least 10 minutes each rather than a small set of panel talks.
    \item Information on what programs are available for future funding
    \item Network with other CI-minded researchers; Get ideas for improving our program; Learn about other projects and methods funded by the CyberTraining program
    \item I want to interact with the community and learn about other PIs' accomplishments. 
    \item I would also like to know about upcoming funding opportunities and ways to write successful sustainability and CyberTraining proposals.
    \item networking, making products visible, offering best practices in training
    \item Insight as to how to write a successful CyberTraining proposal
    \item Networking and collaborations
    \item I don't know
    \item Learn about existing and future CyberTraining and CSSI project scopes and prepare future submissions.
    \item best practices for increasing diverse participation and sustainability of programs
    \item Obtain valuable experiences and lessons in NSF CSSI and CyberTraining Implementation
    \item Learn from and communicate with the other projects and look for collaboration opportunities.
    \item Sustainability, best practices, platforms used for training, physical and virtual labs, scaling platforms for CyberTraining
    \item Information about what current CyberTraining programs have to offer for my workforce development activities and possible synergistic activities with CSSI programs
    \item Networking with others who have similar projects.
    \item Identify what succeeded and what was less successful.
    \item Connections with other groups who are doing similar work. Nurture future collaborators. To learn about challenges that others have faced in getting their projects off the ground (this is my first major project) and best practices. Share ideas for CyberTraining best practices.
    \item Networking and learning about other currently active projects
\end{itemize}
\vspace{12pt}
\noindent \textbf{Other topics for sessions}
\begin{itemize}
    \item Multi-layer modeling with adaptive ML/AI community needs of CI
    \item Software systems (besides ChatGPT) for CyberTraining, Comparison of alternative mechanisms for CyberTraining (e.g., GUI vs command line)
    \item -1n-traditional/unusual career paths for training
    \item Strategies for recruiting and retaining underrepresented undergraduate students
    \item upcoming funding opportunities and how to develop successful cyberinfrastructure
    \item AI and Machine Learning education for undergraduates
    \item Best practices in CyberTraining, platforms used for CyberTraining, scaling platforms to expand CyberTraining offerings
    \item who has successfully reached community colleges or MSIs. 
    \item What kinds of CyberTraining programs have been most effective? Sustainability beyond the grant: 
    \item How to motivate institutional support - what has worked and what has not? 
    \item What are the major gaps for DEI and broadening participation. Namely, this is something that we all think about and want to do, but in terms of really being able to do it - where is there a gap between funding and the work that needs to be done to improve these outcomes? 
\end{itemize}
\vspace{12pt}
\noindent \textbf{Comments on Logistics}
\begin{itemize}
    \item Either Box or Google would work
    \item Slide deck might be helpful
    \item online repository with DOI
    \item Would like to obtain a participant list (names, organizations, emails)  to facilitate connections
\end{itemize}

\subsubsection{Post-workshop Survey}
\label{sec:apx-post-survey} 

\paragraph{Survey Stats}
\label{sec:apx-post-surv-stats}
In addition to the pre-meeting survey described above, we conducted a post-meeting survey was sent out to 84 participants; 30 responded, giving a response rate of 36.6\% recorded in Table \ref{tab:post-surv-resp-stats}. The survey is located online \cite{PostSurveyForm}. The survey was designed to be simple and fast, so it only had eight questions (see below), most of which were selection matrices. The survey questions asked participants to rate the usefulness and satisfaction of the program and meeting logistics using a Likert scale of 1:5 (see Table \ref{tab:likert2}. All survey response data are shown in the Tables \ref{tab:post-surv-resp-stats}, \ref{tab:post-surv-usefullness}, \ref{tab:post-surv-satisfaction}, and \ref{tab:post-surv-cssi}.
\begin{table}[h!]
\vspace{6pt}
\caption{CyberTraining PI Meeting: Survey Response Data}
\noindent\makebox[\textwidth]{
\begin{tabular}{p{2.5in}  p{1.5in} p{0.75in} } 
\toprule
RESPONSE DATA & VALUE \\
\hline
Total emails sent (\#registered): & 82 \\
Number responded: & 30 \\
Response Rate & 36.6\%  \\
\bottomrule
\end{tabular}
}
\label{tab:post-surv-resp-stats}
\end{table}

\paragraph{Meeting Usefulness \& Satisfaction}
\label{sec:apx-post-surv-usefulness}

Table \ref{tab:post-surv-usefullness} summarizes respondents' opinions of the usefulness of the sessions. The overall usefulness of the meeting was 4.5, indicating that attendees enjoyed the meeting and found it beneficial. For the different sessions, all received over 4.3+ average ratings. The poster (4.8/5.0) and NSF (4.7/5.0) participations were the most popular, while the lightning rounds were the least popular (4.23). The lightning rounds were challenging for two reasons: (1) they were held during breakfast, which could have been an opportunity to network with other PIs; and (2) there were so many PIs participating that it was challenging to get them all done in the time allotted. 
The overall satisfaction for the meeting was a bit lower, 4.1, and this is mainly due to the location and hotel. In spite of this, the participants were overall vwery satisfied.

The tables showing the data for the Meeting Usefulness and Participant Satisfaction survey data are shown in the same name tables below. The overall usefulness of the meeting was 4.5, indicating that attendees enjoyed the meeting and found it beneficial. For the different sessions, all received over 4.3+ average ratings. The poster and NSF participation were the most popular (about 4.7/5.0), while the lightning rounds were the least popular (4.23). The overall satisfaction for the meeting was a bit lower, 4.1, and this is mainly due to the location and hotel. In spite of this, the participants were overall satisfied. 

\begin{table}[h!]

\vspace{6pt}
\caption{CyberTraining PI Meeting: How useful did you find the following?}
\noindent\makebox[\textwidth]{
\begin{tabular}{p{2.5in}  p{1.5in} p{0.75in} } 
\toprule
TOPIC & RESPONSE \\
\hline
Poster Sessions & 4.8  \\
Shared Sessions & 4.4  \\
NSF Participation & 4.7  \\
Breakout Sessions &  4.30  \\
Lightning Talks &   4.23  \\
\hline
Overall average usefulness of the meeting & 4.5  \\
\bottomrule
\end{tabular}
}
\label{tab:post-surv-usefullness}
\end{table}
\begin{table}[h!]

\vspace{6pt}
\caption{CyberTraining PI Meeting: How satisfied were you with the following?}
\noindent\makebox[\textwidth]{
\begin{tabular}{p{2.75in}  p{1.0in} p{0.75in} } 
\toprule
TOPIC & RESPONSE \\
\hline
Location of the venue & 3.77 \\
The hotel & 4.10  \\
Meals & 4.27  \\
Your overall satisfaction & 4.27  \\
\hline
Respondents overall average satisfaction & 4.1  \\
\bottomrule
\end{tabular}
}
\label{tab:post-surv-satisfaction}
\end{table}

\paragraph{CSSI Colocation}
\label{sec:apx-post-surv-cssi-colo}
\noindent In response to whether or not participants enjoyed the co-location of the meeting with CSSI, the response was overwhelmingly yes (28/30), and 2 neutral (see Table \ref{tab:post-surv-cssi}. In addition, three respondents indicated an interest in helping with next year’s meeting.  In conclusion, the survey indicates that the workshop was successful, and we should plan to co-locate with CSSI in the future.

\begin{table}[h!]

\vspace{6pt}
\caption{CyberTraining PI Meeting: CSSI Co-location Responses }
\noindent\makebox[\textwidth]{
\begin{tabular}{p{3.0in}  p{1.in} p{0.5in} } 
\toprule
QUESTION & RESPONSE & AVG \\
\hline
Did you enjoy co-locating the meeting with CSSI? & 28/30 & 93\% \\
Did you make any new connections at the meeting? & 27/30 & 90\% \\
Should we co-locate with CSSI next time? & 26/30 & 87\% \\
\bottomrule
\end{tabular}
}
\label{tab:post-surv-cssi}
\end{table}

\paragraph{Experience questions}
\label{sec:apx-post-surv-exper}

\noindent \noindent As part of the survey, we asked three experience questions, requiring text input. The number of responses ranged from 18 to 29, indicating that there is significant interest in the meeting and that participants benefitted from attending. 
\begin{itemize}
    \item Q5-Please name something you learned that will help you in your work  (29 responses) 
    \item Q6-What did you particularly enjoy about the meeting (21 responses)
    \item Q7-What suggestions do you have that would help us create a better overall experience next time? (18 responses)
\end{itemize}

The last question (Q7) had 29 responses, and most of the input was positive. However, there were a few references indicating that: lightning talks were not popular and could have been better organized; the hotel was not popular; and we should start planning earlier. One response, from Q5, summarizes the type of experience we were hoping to provide: 
\begin{quote}
\textit{The workshop provided an invaluable opportunity for me to learn from other projects. I was able to connect with researchers from entirely different domains and discuss potential future collaborations. As a grant awardee, I gained significant insights into the CyberTraining program. Overall, the workshop was a truly enriching experience.}
\end{quote}

The question responses are summarized below.
\begin{itemize}
    \item \textbf{Q5: Please name something you learned that will help you in your work.}
\begin{itemize}
    \item Better understanding of NSF metrics
    \item Better vision of the project from previous successful stories different training approaches
    \item Funding directions and priorities
    \item Getting overview of the CyberTraining program is very useful. Also useful to see other's projects and talk to PI
    \item GPU is extremely important and expensive. How to efficiently share GPUs will be a problem.
    \item How the different NSF CI programs fit together.
    \item How to carry out the CyberTraining project and the position of CSSI projects.
    \item How to promote participation from underrepresented communities; how to evaluate the training.
    \item how to write a good NSF proposal for CyberTraining
    \item I got an idea for how to pitch sustainability
    \item I learned about other new modules, developed by other funded projects.
    \item I learned about some ideas for peer-mentorship and additional resources to learn more. I also learned about other groups educational materials that would be worth incorporating into my own project to train students.
    \item Informatics Skunkworks Program
    \item Knowing the landscape of other CyberTraining projects, their practices, and their work.
    \item Learned about several related CyberTraining projects.
    \item Learned about some similar projects and their approaches, also connected with others interested in our approach/results.
    \item Meeting fellow awardees and hearing priorities was useful, and helps to better understand how a collective impact strategy might be implemented
    \item New resources
    \item not sure
    \item People doing similar things to leverage so we don't duplicate effort strategies for keeping CyberTraining participants (the learners) engaged in a long-term program
    \item Sustainability
    \item The activities of the funding projects inspired me.
    \item The increased use of AI in education and thus CyberTraining.
    \item The spectrum of activities (and a set of common denominator items) that people are pursuing
    \item The workshop provided an invaluable opportunity for me to learn from other projects. I was able to connect with researchers from entirely different domains and discuss potential future collaborations. As a grant awardee, I gained significant insights into the CyberTraining program. Overall, the workshop was a truly enriching experience.
    \item There is a great need for federated learning materials repositories
    \item Topics covered by current CyberTraining grants
\end{itemize} 
\vspace{6pt}
\item \textbf{Q6: What did you particularly enjoy about the meeting?}
\begin{itemize}
    \item Being able to network with other researchers.
    \item The Broadening participation panel was great; I loved the co-location aspect, and that should definitely be repeated.
    \item Chatting with the other PIs and learning about their projects (challenges, strategies for overcoming, etc.)
    \item Connecting with people outside my field.
    \item Discussing potential interactions with colleagues and learning in person what they are doing
    \item Discussions
    \item During the breakout, discuss successes before discussing challenges.
    \item I enjoyed hearing about others' successes and challenges that mirrored experiences that I have had.
    \item I enjoyed talking with other CyberTraining people
    \item lightning talks and poster sessions
    \item Meeting new colleagues working in similar areas
    \item meeting people
    \item Meeting up with fellow CyberTraining folks. Never did this before.
    \item Networking
    \item networking
    \item Networking
    \item Panel discussions.
    \item Poster sessions
    \item Science.
    \item The breakout sessions and the posters.
    \item The plethora of new ideas from breakout sessions.
\end{itemize}
\vspace{6pt}
\item \textbf{Q7:  What suggestions do you have that would help us create a better overall experience next time?}
\begin{itemize}
    \item a venue that is closer to the airport to decrease travel time and costs.
    \item As a first-time attendee, I found the organization and understanding of organization challenging.
    \item Choose a hotel that is closer to an airport
    \item I think the only thing that stood out with a clear suggestion would be having that more strict timing format, or perhaps more clear instructions for lightning talks (or other timed talks) in the future. I think the instructions said 1-slide, but unless people were paying close attention to the talk, number of talks and the time slot. I don't think it was clear initially that they really needed to be 1 minute talks as well.
    \item Managing the time of the Lightning Talks was key. Much better on Day 2. This would have been improved had the PIs been informed in advance that Lightning Talks were limited to 1 minute. More information and clearer communication before arrival would be great. I knew little beyond where to show up (i.e., the hotel).
    \item The meeting venue was too cold!
    \item Maybe find a hotel closer to the airport
    \item More time for breaks
    \item My grant is over with no funds left. It was strange to be told this meeting was required and then not offered the possibility of having my travel covered. As a result, I ended up covering some of the costs out of pocket.
    \item NA
    \item None. Excellent!
    \item On the hotel side: (1) if the rooms cannot be that cold, it will be great. (2) Special needs meals need to be respected. They did not care for gluten-free and dairy-free needs. The hotel location is PERFECT. In a nice not-so-downtown area, and safe. Hotel rooms met our needs, so I can't complain.
    \item Remove the lightning talks
    \item Start planning the meetings earlier
    \item Such meetings should be designed around maximizing N-to-M personal interactions. If co-located with CSSI again, the two meetings should actually be fully jointly organized, not just co-located with a few overlaps.
    \item The location and hotel were subpar (many people had issues with the plumbing, and lack of hot water); the hotel quality aspect couldn't necessarily be known in advance.
    \item Thematic groups
    \item There was a lot of interaction between CSSI and CyberTraining, but that reduced interactions amongst CyberTraining PIs.
    \item Venue could be closer to an airport.
\end{itemize}
\end{itemize}

\section{Breakout Session Data \& Details}

\subsection{Day 1 Joint Session: Opening Remarks, NSF CSSI and CyberTraining}
\label{sec:day1-welcome}

\begin{itemize}[label={}]
  \item \textbf{Chairs:} \textit{Christine Kirkpatrick (UCSD), Geoffrey Fox, UVa} 
\end{itemize}
\vspace{12pt}

Opening Remarks:	NSF CSSI and CyberTraining kirkpatrick.pdf \\
Session Chairs:  	Christine Kirkpatrick, University of California, San Diego
	Geoffrey Fox, University of Virginia \\
Welcome:  	Melissa Cragin, Rice University \\
NSF Speakers:	\\
* Varun Chandola, NSF Cyberinfrastructure for Sustained Scientific \\
* Innovation (CSSI) [chandola.pdf] \\
* Ashok Srinivasan, NSF CyberTraining [srinivasan.pdf] \\
* Chaitan Baru, NSF Directorate for Technology, Innovation, and Partnerships (TIP) [baru.pdf] \\

Joint sessions were hosted in the main conference room, where meals were also served. As a result, meals also served as opportunities for both CSSI and CyberTraining projects to interact.

\subsection{B1A: Best practices in CyberTraining lessons from projects Breakout Session}
\label{sec:appdx-b1a}
\begin{itemize}[label={}]
    \item \textbf{Session Chair: } \textit{Dirk Colbry, MSU} 
    \item \textbf{Session CoChair:} \textit{Mary Thomas, SDSC}  
    \item \textbf{Session Scribe:} \textit{Zhen Ni, FAU} 
\end{itemize}
\vspace{12pt}

\begin{enumerate}[label=\arabic*., start=1]
    \item What is CyberTraining?
    \begin{enumerate}[label=\roman*.]
        \item The training that trains students/researchers on cyberinfrastructure-related skills and knowledge.
        \item Training to develop research workforce in CI technologies.
        \item Anything includes computer, data acquisition, HPC skills, emergent cyber skills, and so on. Data could be from different sources or disciplines.
        \item Training students/faculty/researchers on CI skills, methods, and platforms to enable them to do research.
        \item Training interested learners in how to use computing systems to perform desired research or tasks.
        \item CyberTraining is the art of teaching people to use HPC and cyberinfrastructure resources.
        \item Workshops to train educators in training future CI people (application researchers, sysadmins, HPC researchers).
        \item Infrastructure (machine, curriculum) to enable others to teach CI.
        \item Spread cyber technologies in communities.
        \item Refer to the NSF CyberTraining program website.
        \item CyberTraining = cyberinfrastructure + Training (who, what, where)
    \end{enumerate}
    \vspace{12pt}
    \item What motivated your current projects?
    \begin{enumerate}[label=\roman*.]
        \item CyberTraining beyond computers. The adoption of machine learning in geoscience. Using computers and tools
        \item A consistent training way to organize workshops, training students, and so on.
        \item I am in the field of marine science. A lot of students in marine science are intimidated by coding, data access, analysis, visualization, HPC, etc.
        \item Need to support AI researchers with CI and the corresponding gap in AI skills in CI Professionals.
        \item The lack of connections between PDC experts and instructors who teach CS.
        \item Train faculty, students, and staff to use our HPC clusters and support their research.
        \item Ensure training opportunities for small groups as well as larger groups with established research software engineer support in dark matter.
        \item The lack of collaboration between experts in designing/reusing/improving educational materials, especially in PDC.
        \item I Need to teach people how to use large-scale systems at my center, and there is also the need for funding to do this.
        \item Filling the gap between aspiring cybersecurity researchers and HPC and CI technologies (AI, ML, Big data, parallel computing...).
        \item Need for funding to train facilitators beyond system admin support.
        \item Train young researchers/students so they can move forward with fewer obstacles, to study satellite images. There is substantial potential to benefit marine science. Students have difficulty in related knowledge, such as data acquisition and data visualization. A lot of training is needed for students to start. Usually, the program relies on students themselves to study themselves.
        \item Avoid bias in DS education. Leverage academia and industry to report on DS education challenges.
        \item There is also a need to share developed curriculum materials and reduce redundancy.
    \end{enumerate}
    \vspace{12pt}
    \item What worked?
    \begin{enumerate}[label=\roman*.]
        \item Students loved the hands-on components of the workshop.
        \item In-person training; pre-training data camp; 
        \item Students loved in-person training (worked!)
        \item Covid/virtual meetings once we learned from Zoom. And virtual will not go away
        \item Virtual, spread across several weeks, all sessions recorded.  Worked for busy CI Professionals.
        \item Recording training materials and keeping them permanently online was successful.
        \item For certain types of courses, virtual workshops can work if we have the appropriate approach and pedagogy.
        \item Pre-workshop activity, especially simple things (like "collect XYZ in one place before you come"). worked
        \item Tying the workshop with other synergistic events, e.g., REU.
        \item Posting recordings of training on our YouTube channel.
        \item Students may like to work with working templates. Otherwise, students may be hesitant to join.
        \item Having student helpers present to those having difficulty with various tasks.
        \item Group study (hackathon project) works.
    \end{enumerate}
    \vspace{12pt}
    \item What didn’t work?
    \begin{enumerate}[label=\roman*.]
        \item Post-surveying participants generated a low response rate.
        \item Diverse skill levels in learners can't please everyone; too easy for some, too hard for others.
        \item Assigning work as part of a Workforce Development training event (webinars, etc) even when GitHub is available
        \item Follow-on skill development past the workshop
        \item Inter-institutional challenges due to the legal framework governing universities. Lots of red tape, parsing through agreements, and working through liabilities incompatibility.
        \item Universities are not cooperative with the "participant support cost" to be paid to students and training participants.
        \item I'll train you now, you'll finish it later
        \item Understanding participant support vs other funding for a project.
        \item Organizing in-person events requires a LOT of work with the institutions: bureaucracy.
        \item Record "NSF style" registrations and participation.
    \end{enumerate}
    \vspace{12pt}
    \item What do you use to evaluate success?
    \begin{enumerate}[label=\roman*.]
        \item Online survey for feedback
        \item Surveys before, during, after, long-term
        \item Metrics: attendees, attrition, surveys, views
        \item Instruments: Survey, quizzes, interviews.
        \item Longitudinal feedback (usually more anecdotal than systemic): observing the course of past trainees, e.g., joining advanced degrees, using the CI techniques…
        \item Hackathon project: the students form teams to work on a project, and give a final presentation.
        \item Use a third-party evaluator.
        \item How long it takes students to make their first plot.
        \item We ask participants to take notes during the workshop and use that to understand how transformative the experience was.
        \item Evaluators in psychology (for instance) understand statistics so much better than we do. Further, they understand how questions prime the people polled and influence their answers.
        \item How many courses have been entered in our system (CS Materials).
        \item If participants send you their colleagues, you have succeeded.
        \item External evaluator who helps form questions and assessment
        \item Publications -- reviews provide good input
        \item Best practice idea: build resources and guide on how to find or work with an evaluator.
        \item How happy participants are with the workshop at the end
        \item Write the NSF report in a way that is publishable in suitable professional conferences such as PEARC, SCxx, etc.
        \item Create a list of programs and journals for publications
    \end{enumerate}
    \vspace{12pt}
    \item What do you need to know to write a CyberTraining grant?
    \begin{enumerate}[label=\roman*.]
        \item Local institutional policy on paying participants, organizing workshops, including paying for food and housing
        \item Sustainability: how I will be funded past the NSF grant.
        \item Have a plan for communicating your activities through the community (email does not have enough impact).
        \item Have a clear target population
        \item Make sure the program is scalable and sustainable
        \item Research what has already been funded in your program or related programs and identify an area of need
        \item How to find an evaluator
        \item Solve the platform needs before submission or be the first 6 months.
        \item Know your target learners. Design your program to meet their needs.
        \item How to keep it long-standing, and how to scale up. How to train the trainer who will train more people.
        \item How is it going to scale? Is there an exponential impact to it? => Train-the-trainer is appealing to NSF.
        \item Partnership with the university's professional development / continuing education unit.
        \item Partner with industry to adopt the training program.
        \item Research what has already been funded in your program or related programs and identify an area of need
        \item Paying for the training program out-of-pocket? People would be willing to pay to come to the training.
    \end{enumerate}
\end{enumerate}

\subsection{B1B: Approaches to broader impact and participation: Breakout Session and joint panel with CSSI}
\label{sec:appdx-b1b}

\begin{itemize}[label={}]
    \item  \textbf{Session Chair:}   \textit{Jeff Carver, Alabama}
    \item  \textbf{Session CoChairs:}   \textit{Irfan Ahmed, Virginia Commenwealth, Nicole Gasparini, Tulane}
    \item  \textbf{Session Scribe:}   \textit{Marisa Brazil, ASU}
\end{itemize}

\vspace{12pt}
\noindent This session was organized into three breakout sessions: \\
\begin{enumerate}
    \item Hope: Working with Community Colleges (CC)
    \item Hope: Increase numbers of women (and minoritized scientists?) in computing
    \item Other challenges, inspirations, and observations
\end{enumerate}

\vspace{12pt}
Each session answered or addressed the questions: Why? What? Challenges, and Best Practices, 

\subsubsection{Hope: Working with Community Colleges (CC)}
\begin{itemize}
    \item Why: 
    \begin{itemize}
        \item backgrounds of students very diverse
        \item The first two years are very important, and require that students are helped
     \item Community college students may  be “invisible” to the job market/employers - they don’t qualify due to requirements.
    \end{itemize}
    \item What:
    \begin{itemize}
        \item aligning cc students with industry and providing career pathways
        \item  problem-based learning to solve challenges
        \item focus on tools vs theory (CC’s vs. 4-year colleges) - gap
    \end{itemize}
    \item Challenges:
        \begin{itemize}
        \item CC students might have competing priorities
        \item lose the instructors to industry jobs
        \item What are the best practices for interacting with diverse communities?
        \item How do we make connections with community colleges? Just sending an email to a few professors at an HBCU or a CC doesn't work. If you are lucky enough to be at a university that has people who build those relationships, you have a higher chance of success.
        \item even if the class is offered, will they take them?
    \end{itemize}
    \item Best Practices:
        \begin{itemize}
        \item 
        \item  working with CC instructors to develop tutorials
        \item  reach out to the department chairs at the community colleges to identify needs
        \item include instructor in community college in the proposal development
        \item Requires long-term partnership - sustainability
        \item Offer Virtual Machines or access via the Cloud 
    \end{itemize}
\end{itemize}

\subsubsection{Hope: Increase numbers of women (and minoritized scientists?) in computing}
\begin{itemize}
    \item Best Practices:
        \begin{itemize}
    \item Need to introduce students to stem earlier in their education (elementary)
    \item Role models
    \item  Summer Camps
    \item Engage parents 
    \item  Bilingual programs 
    \item  Make sure there is an inclusive environment 
    \item  Make sure that instructors for early courses are diverse
    \item Engage students in out-of-classroom learning, experiential learning
    \item Creating a learning environment and inclusive activities
    \end{itemize}
    \item Example programs:
    \begin{itemize}
    \item AI for All
    \item Mira! Sparqs program at NAU
    \item ASU Labriola - safe and inclusive space for students from Indigenous communities (other programs are available, too)
    \end{itemize}
    \item Challenges:
    \begin{itemize}
    \item How do we retain students that are already there? 
    \item When broadening participation, the metrics may not necessarily be in large numbers, but in sustainability impacts
    \end{itemize}
\end{itemize}

\subsubsection{Other challenges, inspirations and observations}
\begin{itemize}
    \item Challenges:
    \begin{itemize}
    \item At small universities, how do you introduce CyberTraining without large numbers of faculty?
    \item How do we incentivize faculty and students to participate and be involved in broadening teaching offerings?
    \item How do we organize community exchange/partnership?
    \item How do we get support for pre-tenure faculty who participate in this? They have a lot of things pulling at them, and will this be seen as valuable to participate?
    \item What are strategies for recruiting and retaining underrepresented undergraduate and graduate students?
    \item How do we reach out to the populations that are not our own?
    \item Challenges aren't just teaching broad faculty to use CyberTraining modules. They also need time to integrate tools into their classes. If they don't have the resources (money/time) to integrate it, it's a waste.
    \item  The budget to pay participants is important.
    \item Train the trainers is a great idea, except where is the money supposed to come from to continue the proliferation of training beyond the life of the grant?
    \item How do we interface with universities that we don't usually work with? Are there companies/can NSF that enable this?
    \end{itemize}
    \item Example programs:
    \begin{itemize}
        \item Physics for Poets
     \item Data literacy for everyone - data is for everyone (STEM and non-STEM) - make this part of the core curriculum - champion data literacy for everyone (IS THIS AN EXISTING PROGRAM?)
   \end{itemize}
   \item Best practices:
    \begin{itemize}
    \item Early courses - everyone takes them, and specialized courses are offered later
    \item instructor training (at CC's/other?), stipends, teaching them, and they can take back to students, partnership, develop and test in the classroom
    \end{itemize}
    \end{itemize}

\subsection{B1C: Technology for CyberTraining, including ChatGPT and AI Breakout Session}
\label{sec:appdx-cytr-tech}

\begin{itemize}[label={}]
        \item  \textbf{Session Chair:}   \textit{Venkatesh Merwade, Purdue}
        \item  \textbf{Session CoChairs:}   \textit{Prasun Dewan UNC Chapel Hill, Jessica Eisma, UTA}
    \item  \textbf{Session Scribe:}   \textit{Henri Casanova, Hawaii}
    
\end{itemize}
\vspace{12pt}

\noindent This breakout session primarily focused on four topics: (i) types of technologies used in CyberTraining; (ii) alternative systems or mechanisms used in CyberTraining: (iii) challenges associated with the use of technology in CyberTraining: and (iv) role of ChatGPT/AI in CyberTraining. Discussion from each topic is summarized below.

\vspace{12pt}
\noindent \textbf{Technologies used for CyberTraining:} \\
There is a range of technologies used for cyber training, including Jupyter Notebook, python programming, Docker/containers, Google Colab, Eclipse, Raspberry-pi, GitHub, myGeohub, Sk-Learn, Zoom, Flask, and YouTube videos. The use of the Tapis framework for sensor data, DataCamp for data science, and the use of IBM Cloud for logging user actions was highlighted. Additionally, some members also highlighted the role of a cost-effective in-house cloud featuring bare-metal servers and VM hypervisors, offering customization and hardware attachment capabilities. Python and Jupyter Notebook emerged as the most commonly used programming language and platform, respectively, among most participants in the breakout group, as shown in the word cloud below.
 
\vspace{12pt}
\noindent \textbf{Alternative Mechanisms for CyberTraining:} \\
The alternative mechanism here means something different than the norm. For example, the use of visualization to make an inference instead of quantitative analysis. Multiple alternative mechanisms and their uses were discussed. Most participants preferred graphical user interfaces (GUIs) for lower-level courses, while command line is being used for advanced students. Customization is preferred to cater to different user needs. Open-source tools are used, including Raspberry Pi, Jupyter Notebooks, and RDIs, primarily for hardware training. Visualization and machine learning tools like SAGE 3 are used to monitor progress, often through Zoom, YouTube, and GitHub. Some participants use and/or suggest peer mentoring by senior students who receive credits to help junior students with tools like GitHub and Python. Several participants suggested data science training through boot camps, mentoring programs, and research symposiums, emphasizing hands-on learning and peer teaching.  Workflow-based training was also highlighted.

\vspace{12pt}
\noindent \textbf{Challenges:} \\
Two primary challenges emerged from this discussion, including scaling and access. There's a challenge in scaling CyberTraining platforms to accommodate a large number of participants. The need for a reliable, open platform for content delivery was discussed, as commercial platforms are seen as too commercial. However, some participants mention using commercial platforms for scaling. Several participants mentioned the need and use of high performance computing and distributed computing resources for CyberTraining. Accessing NSF resources such as ACCESS \cite{ACCESS} can be challenging for non-computer science students, but containers can assist.

\vspace{12pt}
\noindent \textbf{Role of ChatGPT/AI:}  \\
AI, including ChatGPT, is used for tasks such as extracting data from research papers, prompt engineering, and training. It's employed as a tool to facilitate learning by asking questions and checking understanding. The discussion acknowledged that questions about the effectiveness of AI in education remain, and ChatGPT is not used as the sole definitive source of information. Efforts are being made to integrate generative AI and large language models like ChatGPT into courses at different levels of AI integration, treating them as tools to expedite processes. Ethical considerations were raised regarding the use of AI, including concerns about corporate data being put into ChatGPT and the need for guidelines to determine when ChatGPT usage may become unethical. Despite its limitations, ChatGPT is seen as a valuable tool for improving writing style, finding programming bugs, and providing support when learning new skills or languages. Examples were given of how students have benefited from interacting with ChatGPT, such as understanding concepts better through repeated questioning and using logs of prompts to identify areas where students struggle. The anonymous nature of interacting with ChatGPT encourages students to ask more questions without feeling self-conscious.

Overall, the discussion highlighted both the potential and limitations of ChatGPT in education and the importance of ethical considerations in its use.

\vspace{12pt}
\noindent \textbf{CyberTraining Issues raised by Relevant Technologies.}  \\
Based on the discussion above, we see four kinds of technologies relevant to CyberTraining:
\begin{enumerate}
    \item The programming environment trainees use for hands-on training, such as Jupyter/Python, the command line, and workflow systems.  
    \item Technologies for visualizing the progress of trainees. Such visualization can allow the trainers to (a) better understand the problems the trainees are facing and (b)  possibly intervene to help them.
    \item Technologies for logging trainee actions. Such logging is needed for visualization.  It is also essential for objective evaluation of the training task.
    \item ChatGPT: This technology can automate some of the trainers' actions by providing intelligent tutoring.
\end{enumerate}
\vspace{12pt}
Common issues raised by each of these technologies are: 
\begin{enumerate}
    \item How does a CyberTraining project get access to and meaningfully use scalable and cost-effective versions of these technologies? For instance, how does a project get access to or create a scalable visualization or logging technology that can incrementally track trainees' progress? How does it make meaningful use of logging/visualization? Similarly, how does a project make meaningful use of the ChatGPT technology ( user interface or its API)  to provide training support? How does a project use the ChatGPT technology to support cost-effective, real-time training?
    \item Training in technology: Assuming the above problems can be addressed, what impact does the use of each of these technologies have on the training given to the trainees? For example, what training has to be given to the trainees to use (a) a nonstandard, or even a standard, programming environment, (b) a technology that allows logging/visualization of their actions, and (c) ChatGPT?
    \item Many of these technologies can be used in multiple projects. How can common techniques and technologies be shared effectively by these projects?
\end{enumerate}
\vspace{12pt}

\subsection{B2A: CyberTraining Metrics and Outcomes Breakout Session}
\label{sec:appdx-cytr-metrics}

\begin{itemize}[label={}]
        \item  \textbf{Session Chair:}   \textit{ Sushil Prasad, UTSA}
        \item  \textbf{Session CoChair:}   \textit{Lonnie Crosby, UTK}
    \item  \textbf{Session Scribe:}   \textit{Prabha Sundaravadivel, UT Tyler} 
\end{itemize}
\vspace{12pt}
Discussion details for Session B2a: Metrics and Outcomes. The session was organized under two main categories:

\vspace{12pt}
\begin{enumerate}
    \item \textbf{B2a: Metrics and Evaluation:}
    \begin{enumerate}
    \item \textbf{Important project metrics:}  Beyond the identification of usual project metrics, such as the number of attendees, some important metrics were identified that illustrated leaky pipelines (registrations, attendees, attendees who fully/actively participated, or attendees who completed assignments).  Some suggestions for metric collection methods include logging keystrokes during examinations, using automatic graders, and real-time assignment status and feedback.
    \item \textbf{Role of surveys:}  Identified several roles of surveys in the metric collection and project evaluation, including the evaluation of satisfaction, identification of improvement opportunities, and the tracking of knowledge before and after training.  Additionally, some challenges of surveys were identified, such as low response rate, lack of motivation, biased responses, and the impact of wording or software choices in survey construction and administration.
    \item \textbf{Role of learning objective-oriented assessment:} Identified tests and exams as a method of evaluation.  The approach of identifying learning objectives may provide a way to scale curriculum based on learning level via the assignment of different objectives.
    \item \textbf{Evaluation metrics for indirect project assessments? Can it go beyond the surveys and exit quizzes?} Identified some evaluation metrics beyond surveys and examinations, including looking at participants’ fields of science and whether they have previously participated.  Additionally, the collection of other metrics may provide insights into the quality of participant engagement, such as the number of participants who attend office hours or are asking questions. Finally, some metrics were identified that attempt to measure future impact, such as the number of participants who pursue STEM or obtain internships, the number of peers who adopt curriculum, and measuring direct in-person conversations, media coverage, and tracking broader impacts five years after the project.
    \end{enumerate}
    \item \textbf{Outcomes and Impact}
    \begin{enumerate}
    \item \textbf{How do we measure the impact?}  Identified several ways in which impact is measured, including the number of lab experiments or tests, surveys, interviews, publications, and attendees.  Scalability can be an issue with interviews, and tracking publications requires acknowledgment or references, allowing the use of citation-tracking tools.  Other impact metrics include an assessment of the time spent in the learning path or time spent on resources after training.  It may be useful to have some project funding after the project ends to evaluate the project's impact.
    \item \textbf{Sharing educational materials digitally and incentives for such sharing:}  Identified challenges in digital sharing and incentives to share, such as the expected adoption of materials, effective sharing mechanisms, and whether others could find these materials.  Additionally, the lack of resources to maintain or sustain materials was viewed as a disincentive to sharing.  However, possible solutions to this challenge may include borrowing mechanisms from the open-source community, identification of intellectual property (i.e., licenses), use of citations (i.e., DOIs or publications), utilizing mechanisms to receive feedback on materials via comments, and utilizing personal connections with material recipients.  Another mechanism of material sharing is to create or utilize a community of practice that organizes co-located conferences within applicable fields.
    \item \textbf{Important Broader impacts or other project Outcomes:} Identified some important broader impacts or other important project outcomes, including the extent of workforce development, the ability of participants to obtain internships or positions, and wider community outreach, including ways in which underrepresented sections of the community were affected.  The ability to keep track of professionals beyond training was identified as a difficulty; however, the use of social networks such as LinkedIn may provide opportunities to stay in touch with past participants.
    \end{enumerate}
\end{enumerate}

\vspace{12pt}
\subsection{B2B: Sustainability Breakout Session}
\label{sec:appdx-sustainability}

\begin{itemize}[label={}]
    \item \textbf{Session Chair:} \textit{ Apan Qasem, Texas State Univ.)} 
    \item \textbf{Session CoChair:} \textit{Karen Tomko, OSU}  
    \item \textbf{Session Scribe:} \textit{Marisa Brazil, ASU}  
\end{itemize}
\vspace{12pt}

\noindent Discussion Notes: The sustainability of CyberTraining Projects is Challenging! The cost of sustaining a program beyond the project date is generally not substantial, but it is difficult to secure that additional support. Sustainability can depend on Institutional Support, Scalability of CyberTraining Efforts, Sharing of Educational Resources, Community Building, and Industry Collaboration. The discussion was broken down into “Success Stories” and “Challenges.” Some key points identified were:
\begin{itemize}
    \item PIs should take advantage of diverse funding sources
    \begin{itemize}
        \item Industry, Foundations, Tiered-funding model, NSF TIP 
    \end{itemize}
    \item PIs need to articulate the value of their project work as well as the impact on a variety of stakeholders
    \begin{itemize}
        \item Provide a value proposition for stakeholders  
        \item Demonstrate the impact of projects and work
        \item What has worked well and what hasn’t \end{itemize}
    \item The structure and format of training programs should be adjusted for sustained scalability
    \begin{itemize}
        \item virtual and asynchronous 
        \item packaged to work out of the box 
        \item PI resource commitment
        \item Modular curricular resources
    \end{itemize}
    \item Many success stories of sharing educational material through various platforms 
    \begin{itemize}
        \item Need to consider duplication of efforts (consolidated, searchable repository of training content) 
        \item Account for the fact the trainees, students in particular, have access to other resources
    \end{itemize}
    \item Training programs should be designed to address the needs of diverse communities, which in turn will help with community-building efforts
    \item Further clarity on NSF solicitation in describing the nature of institutional support that is expected for a particular type of grant 
    \item Credit for activities related to sustaining CyberTraining programs can be included in the tenure and promotion policy documents
    \item One needs a centralized, searchable platform/repository for sharing CyberTraining material, as was done with the NSF ensemble portal \url{https://www.nsf.gov/awardsearch/showAward?AWD_ID=0840597&HistoricalAwards=false} which is old and seemingly defunct. However, it’s possible that the content has been integrated into NSDL (which was the original goal).
\end{itemize}

\subsection{B2C: The Future: Transformative Directions and New Opportunities Including Collaborations Breakout Session}
\label{apdx:transf-data}

\begin{itemize}[label={}]
        \item  \textbf{Session Chair:}   \textit{Vipin Chaudhary, Case Western Reserve }
        \item  \textbf{Session CoChair:}   \textit{Geoffrey Fox, University of Virginia}
    \item  \textbf{Session Scribe:}   \textit{Erik Saule, UNC Charlotte}
\end{itemize}
\vspace{12pt}

\noindent More details from B2C notes andwWrap Up of B2c :

\vspace{12pt}
\noindent \textbf{B2c: New CyberTraining Focus Areas}
\begin{itemize}
	\item CyberInfrastructure is more and more geared towards AI/ML
	\begin{itemize}
		\item Should CT have a larger focus on these?
		\item  Students are “all” interested in AI.
	\end{itemize}
	\item What are the next big technology gaps, and should CyberTraining foresee them?
	\begin{itemize}
		\item Maybe Quantum Computing?
		\item Net zero
		\item Foundation models in AI.
	\end{itemize}
	\item Should CyberTraining expand to K-12?
	\begin{itemize}
		\item  Ultimately, the K-12 students of today are the CI Pro, CI User, and CI Researcher of tomorrow.
		\item   Need to include teachers (train the trainers)
	\end{itemize}
	\item Should CT seek to fund guidelines for non-CS majors?
	\begin{itemize}
		\item Policy experts and Construction Engineers were given as examples of areas where CyberTraining is important. 
		\item There are many domains being transformed by AI and digital infrastructure.
	\end{itemize}
\end{itemize}

\vspace{12pt}
\noindent \textbf{B2c: Technology for CyberTraining}
\begin{itemize}
	\item Perhaps ACCESS could have a resource designed from scratch to support CyberTraining. Large computational demands of AI have overwhelmed many educational resources.
	\item Google Colab enhanced with technologies like Microsoft DeepSpeed is needed to cover the range of interactive single GPU to multi-GPU jobs. 
	\item Many consider cloud environments, serverless computing, etc., as far more attractive than classic SLURM HPC clusters offered at most universities.
	\item There are AI opportunities in education. 
	\item There seem to be foundational model questions for using AI in education that could help a lot. There are lots of FERPA issues, though. 
	\item There are questions as to which data we can use for AI training. Is data to train AI part of cyberinfrastructure?
	\item In other words, can we have technology that helps how we do the CyberTraining? 
	\item Maybe that adapts to different people at different skill levels, for instance, differential learning. One of the differences with the way this is used in other educational contexts is scale. 
	\item There are millions of high schoolers, but only a few people learning your particular chemistry PDE solver.
\end{itemize}

\vspace{12pt}
\noindent \textbf{B2c: Collaborations}
\begin{itemize}
	\item NSF could facilitate in some way interactions between efforts and industry such as Meta, Intel, Google, NVIDIA, IBM, or applications GE GM Deere …
	\item They already provide training.
	\item  They have the machines.
	\item   They have the data.
	\item CSSI has been developing great tools.
	\begin{itemize}
		\item  Can we identify those CSSI projects that need training content and support from CyberTraining?
		\item  We may want to collaborate with some of the large regional hubs, AI institutes, etc. Maybe NSF can help facilitate the connection through Dear Colleague letters.
		\item  Can we collaborate between core CyberTraining and SCIPE?
	\end{itemize}
\end{itemize}

\vspace{12pt}
\noindent \textbf{B2c: Sustainability and Scalability:}
\begin{itemize}
	\item How do we keep the materials alive past the end of the grant?
	\item How do we scale activities up to include more people?
	\item  CyberTraining can help sustain CSSI.
	\item  Need to understand what exists and if one should redo or update existing
	\item  We come back to this in the center/larger grant discussion
\end{itemize}
	
\vspace{12pt}
\noindent \textbf{B2c: Larger CyberTraining Projects (Centers and Institutes)}
\begin{itemize}
	\item It will take professional experts’ help to maintain training materials over the years.
	\item Individual PIs can’t keep running workshops for years past the award ends.
	\item Talking with industry or, more generally, collaborations as a single PI is difficult because of the power imbalance.
	\item  Accessing machines/infrastructure to run training takes resources, as discussed under Technology above.
	\item   Many programs are replicating outreach (advertising)/communication/administration.
	\item  Can bigger efforts enable a multiplicative impact rather than an additive one?
	\item   Centers have longer and larger grants that allow explicit funding for coordination and long-term outreach.
	\item  Coordination can involve an ideas marketplace.
	\item  The large number of CSSI and CyberTraining projects cries out for more coordination and spreading of ideas outside this PI meeting.
\end{itemize}

\vspace{12pt}
\subsection{Success Stories Panel Session}
\label{sec:appdx-success}

\begin{itemize}[label={}]
  \item \textbf{Chair:}  \textit{Mary Thomas, SDSC} 
  \item \textbf{CoChair:}  \textit{Kate Keahey, ANL } 
  \item \textbf{Scribe:}  \textit{Qianqian Liu, UNC Wilmington}   
\end{itemize}
\vspace{12pt}

\noindent The format of the success stories session was to hear from a group of projects about their work and outcomes. We asked each speaker to present a short talk in “lightning” talk form. We provided them with a set of slides and questions that we asked them to address: 
\begin{enumerate}
	 \item Project Overview
	\item What are the resource requirements of your training and education project? How do you find those resources?  What could be improved?
	 \item Do you share educational digital artifacts (e.g., software, homework assignments, etc.) related to your training and education projects? Could your projects benefit from artifacts produced by others?
	 \item What have been your greatest challenges?
	 \item What are the key ingredients to being successful?
\end{enumerate}

In Section \ref{sec:appdx-success}, Table \ref{tab:sucess-stories-presentations} contains a list of all presenters and a link to their talks.

\begin{table}[ht]

\caption{CyberTraining Success Stories session - presentations}
\vspace{3pt}
\begin{tabular}{p{0.15in}  p{1.20in} p{3.15in} p{0.6in} p{0.5in} } 
\toprule
T\#  &  PRESENTER & TITLE  & AWARD & PDF  \\
\midrule
T1  & Henri Casanova &  Integrating core CI literacy and skills into university curricula via simulation-driven activities & 1923621 & \href{https://drive.google.com/open?id=1yWi1j_KaE_o0AAQBqJmHF-2X-zWKHU_r&usp=drive\_copy}{T1-PDF} \\

T2  & Dirk Colbry & A Professional Development and Certification Program (CCIFTD), and CIP-Professional Skills for CyberAmbassadors  & 2118193, 1730137 & \href{https://drive.google.com/open?id=1WSQ96J5W2FR9CDsQDSB3Bmz-A6Y6eEYR&usp=drive\_copy}{T2-PDF}  \\

T3 & Venkatesh Merwade  & CyberTraining for Open Science in Climate, Water and Environmental Sustainability;
Training for Findable, Accessible, Interoperable, and Reusable (FAIR) science  & 2230092, 2230092 & \href{https://drive.google.com/open?id=1_mcMGazQPew7Lo61fBolSdZ9c5OWDDx9&usp=drive\_copy}{T3-PDF} \\

T4 & Karen Tomko & An Artificial Intelligence Bootcamp for Cyberinfrastructure Professionals  & 2118250 & \href{https://drive.google.com/open?id=1KPXmsmRTCxYNaxJxXEzT8Z59eNR-qs5N&usp=drive\_copy}{T4-PDF} \\

T5 & Eric Shook & Hour of Cyberinfrastructure: Developing Cyber Literacy for Geographic Information Science & 1829708 & \href{https://drive.google.com/open?id=1lnDIzMYIlKUV3g9fbG8aRKp0es7_3Ur7&usp=drive\_copy}{T5-PDF} \\

T6 & Sushil Prasad & Broadening Adoption of Parallel and Distributed Computing in Undergraduate Computer Science and Engineering Curricula; Planning a Sustainable Ecosystem for Incorporating Parallel and Distributed Computing into Undergraduate Education  & 2017590, 2002649  & \href{https://drive.google.com/open?id=189JmizKdSjceduTRP61fn1TL5Yx8YjnU&usp=drive\_copy}{T6-PDF} \\

T7 & Ken-ichi Nomura & Cyber Training on Materials Genome Innovation for Computational Software for Future
Engineers & 2118061 & \href{https://drive.google.com/open?id=1-4un5R9BYCLrJu7jbQOMQyms3GY5lNJY&usp=drive\_copy}{T7-PDF}
  \\
\bottomrule
\end{tabular}
\label{tab:sucess-stories-presentations}
\end{table}

\vspace{12pt}
\textbf{Comments from Audience:}
\begin{itemize}
\item   Best practices to facilitate the adoption of training materials/educational digital artifacts
\item   Whether it is still true that you build it, they will come? Are extra efforts needed to promote the materials?
\item   How do we deal with the short attention span of the learners?
\item  What's your plan to run the project after the money runs out, making the project sustainable?
\item   Possibility to make all the modules publicly available and share all the links. 
\item   For training platforms, what machines are in the back end? 
\item   What modules are available to the public? 
\item   How to participate to get certificates; 
\item   Comment on scalability/interoperability
\item   What are the FACT fellows' (Venkatesh Merwade talk) appointment details? do they access or use external training material?
\item   To Karen Tomko: What are your future plans?
\item   For computing portals, what backend compute resources do you use, and is this a sustainable approach? 
\item   Do undergrad curricula include linear algebra/matrix theory?  How has the popularity of AI impacted HPC/Parallel computing program development and popularity?
\end{itemize}

\vspace{12pt}
\textbf{Questions and Answers for the Panel:} 
Presenters were asked to answer questions as if they were part of a panel. Our scribe notes are below. Some questions stand out:
\begin{itemize}
\item   \textit{Question:} “If you build it (a portal), will they come?” One speaker noted that the average attention span of the average web viewer is 30 seconds, and how can we possibly teach HPC concepts in that short of a time period?  
\item   \textit{Question:}What are long-term funding plans? this raised the question of sustainability.
\item   \textit{Question:}What is the possibility of making all the modules publicly available? NSF has a repository, and there are other efforts out there.
\end{itemize}

\vspace{12pt}
Other questions asked by the audience: \\
\begin{itemize}
\item   \textit{Question:} Best practices to facilitate the adoption of training materials/educational digital artifacts \\
 \textit{Answer:} Other than just putting the digital artifacts online, incentives, for example, extra grant or award, is needed to encourage adoption; We usually need extra committed support to organize workshops, and to see the running products. Potentially, we can look for industry support to facilitate the adoption, but we need to be very careful.  Another way is to target your learners and ask for their needs. If you can address their needs, they will go after your materials. 
\item    \textit{Question}: Whether it is still true that you build it, they will come? Are extra efforts needed to promote the materials? \\
 \textit{Answer:} Not true. A panelist (Dirk) mentioned as a successful program, they have spent a lot of time building and selling the product. It is important to reach out to potential users. You need to let the learners know what you developed meets their needs and can help their research. Be sure to minimize the hurdles for the learners, making the adoption easier for them.
\item   \textit{Question}: How do we deal with the short attention span of the learners? \\
\textit{Answer:} Use three sentences to capture your key points, and then move to three slides, and to 1-hr, or even 3-hr modules. Once the learners see why it is important, they will come back.
\item    \textit{Question}: What's your plan to run the project after the money runs out, making the project sustainable? \\
 \textit{Answer:} The suggestions include Getting institutional support to host the materials. If young faculty members can get credit by maintaining the materials, they are more motivated to do this, go after more funding for long-term support, and ask for a small amount of yearly donation (like \$3) from end users  
\item    \textit{Question}: Is it possible to make all the modules publicly available and share all the links?  \\
\textit{Answer: }There is a repository which includes all NSF projects’ information. Also, there is the new HPC-ED project, which will build a federated repository where you can register information and pull information about other projects.
\end{itemize}

\vspace{12pt}
\subsection{Day 2: Joint Session: Joint Session: NSF CSSI and CyberTraining - Broader Impacts
– Creating and supporting diverse teams and communities}

\textbf{Session Chairs:} \textit{Christine Kirkpatrick, Geoffrey Fox} \\

\begin{table}[ht]
\caption{Day 2: NSF CSSI and CyberTraining - Joint Session}
\begin{tabular}{p{0.78in}  p{1.05in} p{2.8in} p{.5in} } 
\toprule
TIME & PRESENTER & ORGANIZATION  & TALK \\ [0.5ex] 
\midrule
9:00 - 9:15 & Richard Alo & Florida A\&M University & [\href{https://drive.google.com/file/d/1fNh0sDWVR5liwspRR4X75ya-avFAH_Jr/view?usp=drive_link}{PDF}] \\
9:15 - 9:30 & Linda Hayden & Elizabeth City State University, SGX3 & [\href{https://drive.google.com/file/d/1H0Vr4W2g7vmyZavo1lktU63KxN6kvNb2/view}{PDF} ]   \\
9:30 - 9:45 & Dan Negrut & University of Wisconsin-Madison & [\href{https://drive.google.com/file/d/1U_bEHK1tyObnCS38tRhebVMjirijNJXi/view}{PDF}]   \\
9:45 - 10:00 & Sophie Kuchynka & Equity Accelerator & [\href{https://drive.google.com/file/d/1OxuASAvp67ySFcfyTcji2EctHqTO8_sI/view}{PDF}]   \\
10:00 - 10:15 & Reed Milewicz & Sandia National Laboratories & n/a \\
10:15 - 10:30 & Frederi Viens & Rice University & [\href{https://drive.google.com/file/d/1WIy7FAXNMBdRWRIzfvsOG6NV7VSkLcWN/view}{PDF}]  \\
 \bottomrule
\end{tabular}
\end{table}
The importance of broader impact was recognized in our plans, and the topic was covered in a separate joint session with CSSI on the second day. This contained several significant presentations by leaders in the field and three described below applicable to CyberTraining. \\

\noindent \textbf{Speaker Talk Summaries:}

\noindent \textit{Professor Richard Aló} presented on Creating and Supporting Diverse Teams and Communities. He is the Dean of the College of Science and Technology at FAMU and leads the NSF Florida Georgia Louis Stokes Alliance for Minority Participation FGLSAMP. His vision had three components: Reduce the Vast Underrepresentation in STEM (science, technology, engineering, and mathematics), Broaden Participation in STEAM (adding A for Arts), and Work Collaboratively to Equalize the Playing Field.  He gave alarming statistics on the university degrees at different levels granted to different ethnic groups (Asian, White,  HURM - historically under-represented minority). He stressed the value of virtual learning across the geographically broad FGLSAMP community. He described initiatives in data science and cybersecurity to conclude that CI and Data Science Provide Alternatives to Broadening Participation in STEAM. \\

\noindent \textit{Professor Linda Hayden of Elizabeth City State University} covered the results of a study by the NSF Office of Polar Programs Subcommittee on Diversity \& Inclusion. She stressed the uneven state of DEI efforts among different OPP awards. A highlight was the work of the NSF Science and Technology Center CReSIS. The REU program here involved women at the 42\% to 63\% level and minorities at the  66\% to 89\% level over a five-year period. She stressed the importance of partnerships rather than recruitment and recommended working with institutions with an underrepresented focus. So please partner with HBCU (Historically Black College and University), PBI (Predominantly Black Institution), MPO (Minority Professional Organization), or MSI (Minority Serving Institution).  \\

\noindent \textit{Sophie Kuchynka} from the Equity Accelerator studied the role of the mentor-mentee relationship in three scenarios: high school, community colleges, and universities. The value of this was seen in all cases, but near-peer mentorship was particularly effective.

\end{document}